\documentclass[english,aps,prl,twocolumn,superscriptaddress]{revtex4}
\usepackage[T1]{fontenc}
\usepackage[latin9]{inputenc}
\usepackage{amsmath}
\usepackage{graphicx}
\usepackage{esint}

\makeatletter
\@ifundefined{textcolor}{}
{%
 \definecolor{BLACK}{gray}{0}
 \definecolor{WHITE}{gray}{1}
 \definecolor{RED}{rgb}{1,0,0}
 \definecolor{GREEN}{rgb}{0,1,0}
 \definecolor{BLUE}{rgb}{0,0,1}
 \definecolor{CYAN}{cmyk}{1,0,0,0}
 \definecolor{MAGENTA}{cmyk}{0,1,0,0}
 \definecolor{YELLOW}{cmyk}{0,0,1,0}
}

\makeatother

\usepackage{babel}
\begin{document}

\preprint{This line only printed with preprint option}

\title{Electrostatics of solvated systems in periodic boundary conditions}

\author{Oliviero Andreussi}

\email{oliviero.andreussi@unipi.it}

\selectlanguage{english}%

\affiliation{Theory and Simulations of Materials (THEOS) and National Center for
Computational Design and Discovery of Novel Materials (MARVEL), École
Polytechnique Fédérale de Lausanne, Station 12, 1015 Lausanne, Switzerland}

\affiliation{Department of Chemistry, University of Pisa, Via Moruzzi 3, 56124
Pisa, Italy}

\author{Nicola Marzari}

\email{nicola.marzari@epfl.ch}

\selectlanguage{english}%

\affiliation{Theory and Simulations of Materials (THEOS) and National Center for
Computational Design and Discovery of Novel Materials (MARVEL), École
Polytechnique Fédérale de Lausanne, Station 12, 1015 Lausanne, Switzerland}
\begin{abstract}
Continuum solvation methods can provide an accurate and inexpensive
embedding of quantum simulations in liquid or complex dielectric environments.
Notwithstanding a long history and manifold applications to isolated
systems in open boundary conditions, their extension to materials
simulations --- typically entailing periodic-boundary conditions ---
is very recent, and special care is needed to address correctly the
electrostatic terms. We discuss here how periodic-boundary corrections
developed for systems in vacuum should be modified to take into account
solvent effects, using as a general framework the self-consistent
continuum solvation model developed within plane-wave density-functional
theory {[}O. Andreussi et al. J. Chem. Phys. 136, 064102 (2012){]}.
A comprehensive discussion of real-space and reciprocal-space corrective
approaches is presented, together with an assessment of their ability
to remove electrostatic interactions between periodic replicas. Numerical
results for zero-dimensional and two-dimensional charged systems highlight
the effectiveness of the different suggestions, and underline the
importance of a proper treatement of electrostatic interactions in
first-principles studies of charged systems in solution. 
\end{abstract}
\maketitle

\section{introduction}

Computer simulations of materials have been significantly progressing
in recent years due to the many improvements in both computational
tools and underlying algorithms. In particular, density-functional
theory (DFT) has become a very valuable tool to model complex systems
with high accuracy. Even though a large effort in the field has been
devoted to advancing the accuracy of the algorithms beyond the level
of DFT, these improvements usually come with a substantial increase
of the computational costs, therefore imposing some serious limitations
on the system sizes that can be handled. For this reason, hierarchical
algorithms have been developed, which allow to treat different parts
of the systems with different degrees of accuracy, without compromising
the description of the important atomistic features that need to be
characterized. 

Among hierarchical methods, a fundamental role has been played by
continuum dielectric models, which combined with ab-initio and DFT
atomistic calculations have been shown to be very effective in modeling
solvents and complex environments in an inexpensive and accurate way
\cite{tomasi_chemrev_1994,tomasi_chemrev_2005,cramer_chemrev_1999,orozco_chemrev_2000}.
Although most of the continuum dielectric models have been developed
in the chemistry community and applied to study isolated systems,
a large effort has been spent in recent years to extend these models
to the boundary between condensed matter physics and chemistry \cite{fattebert_jcomputchem_2002,fattebert_intjqchem_2003,scherlis_jcp_2006,sanchez_jcp_2009,dziedzic_epl_2011,petrosyan_jpcb_2005,letchworth_prb_2012,gunceler_modmat_2013}.
In particular, the possibility of reducing the computational complexity
of solvated or electrified interfaces would allow the extensive modeling
of a large range of fundamental processes, such as those taking place
in heterogeneous catalysis, electro-chemistry and photochemistry. 

We recently proposed a self-consistent continuum solvation (SCCS)
model \cite{andreussi_jcp_2012,scherlis_jcp_2006,fattebert_jcomputchem_2002}
that combines a highly flexible definition of the dielectric, defined
in terms of a minimal set of parameters, together with an implementation
in a plane-wave pseudo-potential DFT framework that is perfectly suited
to model periodic solid state systems. The model was tested thoroughly
and showed not only an impressive agreement with similar models in
the literature but also very good performance in reproducing the experimental
solvation free energies of neutral compounds \cite{andreussi_jcp_2012}
and charged species \cite{dupont_jcp_2013}. By taking advantage of
fast Fourier transform (FFT) techniques to compute the electrostatic
potential and its gradient in reciprocal space, the overall computational
cost of SCCS is small, and its scaling with system size makes its
impact negligible for large-scale calculations. 

Nonetheless, solving for electrostatic potentials in reciprocal space
is straightforward only when neutral fully-periodic systems are considered.
In the other cases, instead, FFTs approaches can give rise to serious
errors and strong system-size dependence, in vacuum as well as in
a continuum dielectric. In particular, periodic-boundary conditions
are not compatible with charged systems, and so charged systems are
modeled as if they were immersed in a neutralizing charge background
(labeled NCB in the following). Moreover, when periodic-boundary conditions
are used to model heterogeneous or non-uniform systems one needs to
carefully monitor the size of the periodic cell chosen to avoid spurious
interactions with the periodic replicas. For these reasons, it is
well known that simulations of charged systems in the solid state
(e.g. charged defects in semiconductors \cite{shim_prb_2005,schultz_prl_2006,castleton_prb_2006,lany_prb_2008,hine_prb_2009,lany_modmat_2009,freysoldt_prl_2009,freysoldt_pssb_2011,corsetti_prb_2011,komsa_physb_2012,komsa_prb_2012,murphy_prb_2013})
or in explicit solvents solutions (e.g. for solvation energies of
charged ions \cite{hunenberger_jcp_1999,heinz_jcp_2005}), need to
deal with serious artifacts due to the size and periodicity of the
simulation cell. This is particularly important when modeling systems
of reduced dimensionality; nevertheless, the problem is intrinsically
easier to handle than in the three-dimensional case. A wide variety
of approaches has been proposed in the literature to remove the artifacts
due to the presence of fictitious replicas \cite{feibelman_jcp_1984,feibelman_prb_1986,makov_prb_1995,jarvis_prb_1997,martyna_jcp_1999,schultz_prb_1999,minary_jcp_2002,minary_jcp_2004,ismail-beigi_prb_2006,genovese_jcp_2006,genovese_jcp_2007,dabo_prb_2008,dabo_prb_2008_erratum,li_prb_2011,heinz_jcp_2005,hine_prb_2009,hine_jcp_2011,castro_prb_2009}.
One class of methods (labeled here ``non self-consistent'', or NSC)
aims at correcting only the electrostatic energy of the systems, while
keeping the degrees of freedom of the system frozen in the presence
of periodic boundary conditions. This is the approach, e.g., of the
Makov-Payne method \cite{makov_prb_1995}, that is one of the most
widespread methodologies to take care of PBC errors for 0D systems. 

In order to fully remove the effects of periodic boundary conditions
on partially periodic systems, other approaches (labeled as ``self-consistent'',
or SC, in the following) have been developed that correct the electrostatic
potential. This correction enters directly into the electrostatic
energy, Kohn-Sham potential and inter-atomic forces, such that the
electrostatic energy has no spurious contributions from the periodic
replica, but also all the degrees of freedom of the system are optimized
in the correct electrostatic environment. These fully self-consistent
correction schemes can be further divided in two classes, depending
on whether the correction to the electrostatic potential is computed
in real space (R-space)\cite{dabo_prb_2008,dabo_prb_2008_erratum}
or in reciprocal space (G-space) \cite{martyna_jcp_1999,minary_jcp_2002,minary_jcp_2004,li_prb_2011}.
For both classes, correction for 2D, 1D and 0D systems have been proposed
and implemented.

In the present work, some of the existing PBC correction schemes developed
for partially-periodic systems in vacuum are extended in order to
take into account the presence of a continuum dielectric medium in
the system. In the following, the three general classes of corrections,
i.e. NSC, SC R-space, and SC G-space, are analyzed and the modifications
of the algorithms needed to include a continuum dielectric are outlined.
Equations for the most important cases are derived and the proposed
approaches are implemented and tested. 

The paper is organized as follows: in Section IIA we introduce the
notation and the main electrostatic equations used throughout the
article; in Section IIB we review the main equations describing electrostatic
interactions in periodic systems, highlighting the limitations of
standard approaches; in Section IIC we summarize the equations behind
the SCCS model, as derived in Ref. \cite{andreussi_jcp_2012}, underlining
the effects of periodic boundary conditions; in Section IID we describe
the Makov-Payne approach \cite{makov_prb_1995} (NSC, 0D) and appropriately
modify it in order to combine it with the SCCS model; in Section IIE
the density counter charge (DCC) correction scheme \cite{dabo_prb_2008,dabo_prb_2008_erratum}
is analyzed and extended to take into account of the complex dielectric
environment, and its application to the case of slab geometries is
presented (SC, R-space, 2D); in Section IIF the Martyna-Tuckerman
method \cite{martyna_jcp_1999} is discussed and its modifications
are derived and implemented for the case of isolated systems (SC,
G-space, 0D); in Section III we present detailed numerical results
for the 0D and 2D cases; eventually, in Section IV we draw our conclusions.

\section{methods}

\subsection{Electrostatics in periodic boundary conditions}

In oder to establish a consistent notation, we report here the main
electrostatic equations, as reported in many standard textbooks but
with a specific focus on their form in periodic systems. Electrostatic
interactions are governed by Maxwell's equations, which relate electric
field $\mathbf{E}\left(\mathbf{r}\right)$ and charge density $\rho\left(\mathbf{r}\right)$
\begin{align}
\nabla\cdot\mathbf{E}\left(\mathbf{r}\right) & =4\pi\rho\left(\mathbf{r}\right)\label{eq:maxwell1}\\
\nabla\times\mathbf{E}\left(\mathbf{r}\right) & =0.\label{eq:maxwell2}
\end{align}
Due to the irrotational nature of the electrostatic field, it is often
convenient to express it in terms of the gradient of a scalar potential,
i.e. the electrostatic potential, as 
\begin{equation}
\mathbf{E}\left(\mathbf{r}\right)=-\nabla v\left(\mathbf{r}\right).\label{eq:potential}
\end{equation}
and Eqs. \eqref{eq:maxwell1} and \eqref{eq:maxwell2} are recast
into a single second order differential equation, i.e. the Poisson
equation
\begin{equation}
\nabla^{2}v\left(\mathbf{r}\right)=-4\pi\rho\left(\mathbf{r}\right).\label{eq:poisson_vacuum}
\end{equation}

Once a proper set of boundary conditions is imposed, the above differential
equation can be solved exactly. In particular, in a closed volume
of space it is sufficient to specify the potential (Dirichlet boundary
conditions) or the normal component of the field (von Neumann boundary
conditions) at the boundary in order to have a unique solution of
the electrostatic problem. Also, it is customary to recast Eq. \eqref{eq:poisson_vacuum}
in an integral formulation by the use of Green's functions, namely,
\begin{equation}
v\left[\rho\right]\left(\mathbf{r}\right)\equiv v\left(\mathbf{r}\right)=\int_{B}G\left(\mathbf{r}-\mathbf{r'}\right)\rho\left(\mathbf{r}'\right)\mbox{d}\mathbf{r'}\label{eq:potential_green}
\end{equation}
where the integration is performed over the arbitrary bounded region
$B$. In the above equation and in the following, we decided to make
explicit the functional dependence of the potential on the density
that generates it. 

Given the definitions above, the electrostatic energy of a charge
distribution can then be expressed as 
\begin{align}
E\left[\rho\right] & =\frac{1}{8\pi}\int_{B}\left|\mathbf{E}\right|^{2}\mbox{d}\mathbf{r}
\end{align}
For an isolated charge density in vacuum, it is customary to impose
homogeneous Dirichlet or von Neumann conditions at infinity, such
that
\begin{equation}
E\left[\rho\right]=\frac{1}{2}\int_{B}\rho\left(\mathbf{r}\right)v\left[\rho\right]\left(\mathbf{r}\right)\mbox{d}\mathbf{r},\label{eq:energy}
\end{equation}
and 
\begin{equation}
G\left(\mathbf{r}-\mathbf{r'}\right)=\frac{1}{\left|\mathbf{r}-\mathbf{r'}\right|}.\label{eq:green_isol_vacuum}
\end{equation}
For this class of systems, both the potential and the energy can be
easily computed by exploiting Eq. \eqref{eq:green_isol_vacuum} and
by setting the integrand limit in Eq. \eqref{eq:energy} and \eqref{eq:potential_green}
to an arbitrary cell size $D$ large enough to contain the entire
charge density of the system
\begin{align}
v\left[\rho\right]\left(\mathbf{r}\right) & =\int_{D}G\left(\mathbf{r}-\mathbf{r'}\right)\rho\left(\mathbf{r}'\right)\mbox{d}\mathbf{r}'\label{eq:potential_isol}\\
E\left[\rho\right] & =\frac{1}{2}\int_{D}\rho\left(\mathbf{r}\right)v\left[\rho\right]\left(\mathbf{r}\right)\mbox{d}\mathbf{r}.\label{eq:energy_isol}
\end{align}
Nonetheless, the characteristic $1/r$ behavior of the electrostatic
potential can be the source of two specular problems: the divergence
at short distances and the slow decay at large distances make the
electrostatic potential difficult to handle, introducing issues with
the self-interaction of charges and of conditionally convergent calculations
of the field. 

In periodic systems, the fundamental electrostatic equations, e.g.
Eqs. \eqref{eq:potential_isol} and \eqref{eq:energy_isol}, may be
written in the same form reported above, whereas it is intended that
the integration domain corresponds to the periodic unit cell, typically
chosen as the primitive one, and the physical quantities entering
the equations (density, potential, Green's function, etc.) refer to
such infinitely periodic systems. In order to avoid confusion on which
kind of system is considered, in all the equations in the following
sections, we decided to use special typographic characters ($\varrho$,
$\mathtt{E}$, $\mathtt{v}$, $\boldsymbol{\Xi}$, $\mathtt{G}$,
and $\mathtt{D}$) to identify quantities referring to infinite periodic
systems, while keeping the standard labels ($\rho$, $E$, $v$, $\mathbf{E}$,
$G$, and $D$) for localized isolated systems. 

In a periodic system, the entire, infinite, charge density $\varrho\left(\mathbf{r}\right)$
will contribute to the potential $\mathtt{v}\left(\varrho,\mathbf{r}\right)$.
Nonetheless, such a potential can still be expressed univocally with
an integral confined to the unit cell $\mathtt{D}$ of the periodic
system, by exploiting in Eq. \eqref{eq:potential_green} the Green's
function $\mathtt{G}\left(\mathbf{r}-\mathbf{r'}\right)$ appropriate
for periodic boundary conditions
\begin{multline}
\mathtt{v}\left[\varrho\right]\left(\mathbf{r}\right)=\int_{\infty}G\left(\mathbf{r}-\mathbf{r}'\right)\varrho\left(\mathbf{r}'\right)\mbox{d}\mathbf{r}'\\
=\int_{\mathtt{D}}\mathtt{G}\left(\mathbf{r}-\mathbf{r'}\right)\varrho\left(\mathbf{r}'\right)\mbox{d}\mathbf{r}'.\label{eq:potential_pbc}
\end{multline}
Similarly, Eq. \eqref{eq:energy} can also be used as is in order
to compute the electrostatic energy per unit cell of a periodic system
$\mathtt{E}\left[\varrho\right]$, provided that the integration is
over the unit cell $\mathtt{D}$ of the periodic system
\begin{equation}
\mathtt{E}\left[\varrho\right]=\frac{1}{2}\int_{\mathtt{D}}\varrho\left(\mathbf{r}\right)\mathtt{v\left[\varrho\right]}\left(\mathbf{r}\right)\mbox{d}\mathbf{r}.\label{eq:energy_pbc}
\end{equation}

\subsection{Periodic electrostatic potential }

When dealing with periodic systems, it is natural to recast the electrostatic
equations in reciprocal space, in order to exploit the simple form
of the Fourier-transformed differential operator 
\begin{align}
\nabla f\left(\mathbf{r}\right) & \rightarrow\widetilde{\nabla f}\left(\mathbf{k}\right)=i\mathbf{k}\widetilde{f}\left(\mathbf{k}\right)\label{eq:ft_gradient}\\
\nabla\cdot\mathbf{F}\left(\mathbf{r}\right) & \rightarrow\widetilde{\nabla\cdot\mathbf{F}}\left(\mathbf{k}\right)=i\mathbf{k}\cdot\widetilde{\mathbf{F}}\left(\mathbf{k}\right),\label{eq:ft_divergence}
\end{align}
where the overwritten tilde identifies Fourier-transformed functions.
By applying the above relations to Eqs. \eqref{eq:maxwell1} and \eqref{eq:potential},
the general solution of the electrostatic field and potential in a
periodic system can be written as 
\begin{align}
\boldsymbol{\Xi}\left(\mathbf{k}\right) & =-4\pi\frac{i\mathbf{k}\varrho\left(\mathbf{k}\right)}{\left|\mathbf{k}\right|^{2}}\,\mbox{for }\mathbf{k}\neq0.\label{eq:field_k}
\end{align}
and
\begin{equation}
\mathtt{v}\left[\varrho\right]\left(\mathbf{k}\right)=\frac{i\mathbf{k}\cdot\boldsymbol{\Xi}\left(\mathbf{k}\right)}{\left|\mathbf{k}\right|^{2}}=4\pi\frac{\varrho\left(\mathbf{k}\right)}{\left|\mathbf{k}\right|^{2}}\,\mbox{for }\mathbf{k}\neq0.\label{eq:potential_k}
\end{equation}

For $\mathbf{k=0}$ the electrostatic equations need to be handled
with care. Indeed, special forms of the divergence theorem impose
that a periodic solution for the electrostatic field and potential
is only possible provided that the right-hand side of Eqs. \eqref{eq:maxwell1}
and the left-hand side of \eqref{eq:potential}, once transformed
in Fourier space, are zero for $\mathbf{k=0}$. In particular, in
order to obtain a periodic solution for the electrostatic field, the
total charge of the system has to be zero
\begin{equation}
\varrho\left(\mathbf{k}=\mathbf{0}\right)\equiv\left\langle \varrho\right\rangle =\frac{1}{\mathtt{V}}\int_{\mathtt{D}}\varrho\left(\mathbf{r}\right)\mbox{d}\mathbf{r}=0.
\end{equation}
Similarly, a periodic solution of the electrostatic potential will
only be possible for a zero average electrostatic field
\begin{equation}
\boldsymbol{\Xi}\left(\mathbf{k}=\mathbf{0}\right)=\frac{1}{\mathtt{V}}\int_{\mathtt{D}}\mathbf{\boldsymbol{\Xi}}\left(\mathbf{r}\right)\mbox{d}\mathbf{r}=0.
\end{equation}
As this latter condition univocally fixes the constant value of the
electrostatic field, the only undefined quantity for $\mathbf{k}=0$
is the potential: given that the system is neutral, such component
has no effects on the final electrostatic energy
\begin{equation}
\frac{1}{2}\int_{\mathtt{D}}\mathtt{v}\left[\varrho\right]\left(\mathbf{k}=\mathbf{0}\right)\varrho\left(\mathbf{k}=\mathbf{0}\right)\mbox{d}\mathbf{r}=0.\label{eq:energy_k_eq_0}
\end{equation}
Even if $\varrho$ is defined to be non-neutral inside the unit cell,
Eqs. \eqref{eq:field_k} and \eqref{eq:potential_k} can still be
used exactly as written, together with the choice $\mathtt{v}\left(\mathbf{k}=\mathbf{0}\right)=0$,
but the quantities obtained will actually correspond to a periodic
system where the original charge density has been compensated by a
homogeneous background (NCB) 
\begin{equation}
\varrho\rightarrow\varrho-\left\langle \varrho\right\rangle .
\end{equation}
The specific choice $\mathtt{v}\left(\mathbf{k}=\mathbf{0}\right)=0$
is made so that the NCB density does not appear explicitly in the
formulas, since its only contribution to the energy, i.e. the term
for $\mathbf{k}=\mathbf{0},$ cancels out in Eq. \eqref{eq:energy_k_eq_0}.
Nonetheless, for the sake of correctly identifying the physical system
under consideration, in the following we will explicitly write the
dependence of the potential on the compensated charge density of the
system, namely $\mathtt{v}\left[\varrho-\left\langle \varrho\right\rangle \right]\left(\mathbf{k}\right)$.

It has to be noted that the above equations have been derived for
ideally infinite periodic systems, but it could be convenient to take
a different, real-space, perspective and to think of a periodic system
as generated by an increasingly larger number of unit cells. In such
a picture, while the reciprocal space approach can still be used to
look for periodic solutions of the electrostatic field and potential,
it is physically acceptable to have an additional non-periodic, but
linear, component for the electrostatic potential. In other words,
an additional linear potential of the form $\mathbf{\boldsymbol{\Xi}}_{0}\cdot\mathbf{r}$
would still preserve the periodic solution for the electrostatic field,
and thus a physically acceptable solution for the energy of the periodic
system. Moreover, for the same reasons, the $\mathbf{k=0}$ component
of the potential will not have any effect on the total energy of a
neutral system. 

As the $\mathbf{k=0}$ components of the electrostatic field and potential
cannot be univocally determined by the electrostatic differential
equations, they can only be determined by the boundary conditions
imposed on the system. Exploiting Eq. \eqref{eq:potential_pbc}, the
general solution for the electrostatic potential of a periodic system
can be written as 
\begin{multline}
\mathtt{v}\left[\varrho\right]\left(\mathbf{r}\right)=\frac{4\pi}{V}\sum_{\mathbf{k}\neq0}\frac{\varrho\left(\mathbf{k}\right)}{\left|\mathbf{k}\right|^{2}}e^{i\mathbf{k}\cdot\mathbf{r}}+\boldsymbol{\Xi}_{0}\cdot\mathbf{r}+\mathtt{v}_{0},\label{eq:potential_pbc_full}
\end{multline}
where the last two terms are usually referred in the literature as
the extrinsic potential \cite{hunenberger_jcp_1999,heinz_jcp_2005,kastenholz_jcp_2006},
to distinguish them from the intrinsic part, which can be solved independently
of the boundary conditions. In most reciprocal-space approaches to
the electrostatic potential, only the intrinsic part of the potential
is computed, while the extrinsic contributions are assumed to be equal
to zero. This choice corresponds to a specific assumption on the boundary
conditions of the electrostatic problem (spherical surface and tin-foil
boundary conditions, as discussed in the following) and it can introduce
artifacts in periodic calculations of partially-periodic and non-periodic
systems. 

In order to further analyze the expression of the extrinsic contributions,
we can follow the derivation of de Leewen et al. \cite{deleeuw_molphys_1979,deleeuw_prsla_1980a,deleeuw_prsla_1980b,deleeuw_prsla_1983}
and treat the infinite periodic system as a limiting case of a spherical
ensemble of unit cells embedded in a vacuum-like dielectric. Such
a choice univocally determines the electrostatic equations and correspond
to the usual boundary conditions from which Eq. \eqref{eq:green_isol_vacuum}
was derived. Thus the potential can be expressed as 
\begin{multline}
\mathtt{v}\left[\varrho\right]\left(\mathbf{r}\right)=\int_{\infty}G\left(\mathbf{r}-\mathbf{r}'\right)\varrho\left(\mathbf{r}'\right)\mbox{d}\mathbf{r}'\\
=\sum_{\mathbf{R}}\int_{\mathtt{D}}G\left(\mathbf{r}-\mathbf{r}'+\mathbf{R}\right)\varrho\left(\mathbf{r}'+\mathbf{R}\right)\mbox{d}\mathbf{r}'\\
=\int_{\mathtt{D}}\left[\sum_{\mathbf{R}}G\left(\mathbf{r}-\mathbf{r}'+\mathbf{R}\right)\right]\varrho\left(\mathbf{r}'\right)\mbox{d}\mathbf{r}'
\end{multline}
from which, comparing with Eq. \eqref{eq:potential_green}, the periodic
Green's function can be defined 
\begin{multline}
\mathtt{G}\left(\mathbf{r}-\mathbf{r}'\right)=\sum_{\mathbf{R}}G\left(\mathbf{r}-\mathbf{r}'+\mathbf{R}\right)\\
=\sum_{\mathbf{R}}\frac{1}{\left|\mathbf{r}-\mathbf{r}'+\mathbf{R}\right|}
\end{multline}
where the sum over lattice vectors $\mathbf{R}$ is supposed to be
performed over spherical shells around the origin. As thoroughly discussed
by Makov and Payne \cite{makov_prb_1995}, the contribution of the
terms in the periodic sum that determines the electrostatic potential
vanishes as 
\begin{equation}
\frac{q^{(n)}}{l^{n+1}}
\end{equation}
where $q^{(n)}$ is the $n-$th multipole moment of $\varrho\left(\mathbf{r}\right)$
and $l$ is the distance of the shell from the origin. Similarly,
the contribution of each shell of the periodic system to the electrostatic
field in the original cell will vanish as the inverse $n+2$ power
of $l$. For a three-dimensional system, the periodic sum that determines
the potential (field) is divergent for a charge distribution with
non-zero dipole (monopole) moment. This behavior corresponds to the
impossibility, shown above, of obtaining a periodic solution for the
potential (field) in reciprocal space for a system with non-zero electric
field (total charge). Moreover, the periodic sum that determines the
potential (field) is conditionally convergent for a charge distribution
with non-zero quadrupole (dipole) moment, while it is absolutely convergent
for higher multipole moments. Conditional convergence implies that
the results will depend on the order over which the sum is performed
and on the boundary conditions applied. This can be thought as the
result of the fact that a periodic ensemble of quadrupole moments
(dipoles) generates a non-zero surface distribution of dipoles (charges),
which in turns will give rise to a non-zero average electrostatic
potential (field) inside the system. The magnitude of these quantities
will depend on the geometry of the surface of the system and on the
dielectric properties of the embedding medium. For the assumptions
made above (spherical system embedded in vacuum) the expression for
the extrinsic contributions to the potential, first derived by de
Leeuw et al. \cite{deleeuw_prsla_1980b,deleeuw_prsla_1980a}, reads
\begin{align}
\mathbf{\boldsymbol{\Xi}}_{0} & =\frac{4\pi}{3}\frac{1}{\mathtt{V}}\int_{\mathtt{D}}\mathbf{r}\varrho\left(\mathbf{r}\right)\mbox{d}\mathbf{r}\equiv\frac{4\pi}{3}\frac{\mathbf{\mathtt{d}}}{\mathtt{V}}\label{eq:field_ext}\\
v_{0} & =\frac{2\pi}{3}\frac{1}{\mathtt{V}}\int_{\mathtt{D}}\mathbf{r}{}^{2}\varrho\left(\mathbf{r}\right)\mbox{d}\mathbf{r}\equiv\frac{2\pi}{3}\frac{\mathtt{Q}}{\mathtt{V}}.\label{eq:potential_ext}
\end{align}
The above expressions have been recently rederived, for the same system
shape and boundary conditions, by Hunenberger et al. by following
a different approach \cite{hunenberger_jcp_1999,heinz_jcp_2005,kastenholz_jcp_2006}.
In particular, it is important to notice that the constant electric
field that appears in Eq. \eqref{eq:field_ext} is nothing but the
electrostatic field generated by a constant polarization density $\mathbf{P}=\mathbf{d}/V$. 

The extrinsic contributions to the electrostatic potential can be
further extended to the case of a system embedded in a dielectric
medium with arbitrary dielectric permittivity $\epsilon$, while still
keeping the assumption of a spherical geometry. In this case, the
Onsager model of solvation \cite{onsager_jacs_1936} analytically
reduces the effects of the embedding medium to an additional reaction
field that, for the case of a dipolar system, is again constant inside
the system. The classical expression for the Onsager reaction field
\cite{onsager_jacs_1936} gives
\begin{equation}
\mathbf{E}_{R}=-\frac{4\pi}{3}\frac{2\left(\epsilon-1\right)}{2\epsilon+1}\frac{\mathbf{d}}{V},
\end{equation}
which summed to the constant field obtained in vacuum gives the final
result of
\begin{equation}
\boldsymbol{\Xi}_{0}^{\epsilon}=\frac{4\pi}{3}\frac{\mathbf{d}}{\mathtt{V}}-\frac{4\pi}{3}\frac{2\left(\epsilon-1\right)}{2\epsilon+1}\frac{\mathbf{d}}{\mathtt{V}}=\frac{4\pi}{2\epsilon+1}\frac{\mathbf{d}}{\mathtt{V}}.\label{eq:field_ext_diel}
\end{equation}
This expression reduces to the case in vacuum for $\epsilon=1$, while
vanishing when the periodic system is immersed in a perfect conductor
(tin-foil boundary conditions, i.e. $\epsilon=\infty$). 

To summarize, when dealing with the electrostatic equations in periodic
systems two main limitations occur. First, the total charge of the
system needs to be zero, in order to provide a non-diverging solution
for the electrostatic field and the energy of the system. Charged
unit cells can still be treated using Eq. \eqref{eq:potential_pbc_full},
but the potential obtained will be the one of the charge density considered
plus a neutralizing homogeneous background charge (NCB). Second, by
using the standard reciprocal space approach for the calculation of
the potential of a periodic system and by neglecting the extrinsic
contributions to the potential, Eq. \eqref{eq:potential_pbc_full},
a well-defined choice on the boundary conditions of the problem is
made, which can introduce spurious contribution to the energy. 

In addition to the issues alluded to above, the long range decay of
the potential represents a serious drawback for simulations where
periodic boundary conditions are only used as an approximation to
model heterogeneous systems without introducing surface effects. When
studying charged defects in crystals, or solvation energies of ions
and biomolecules in explicit solvents, the electrostatic interactions
coupled with the fictitious periodicity of the cell introduce artifacts
in the simulations that are challenging to handle. 

The problem is even more evident, although easier to solve, when one
wants to model systems of reduced periodicity, being them slabs (2D),
linear systems (1D) or isolated compounds (0D). The problem in these
cases is twofold: first, the electrostatic potential of the ideal
isolated system would not usually show the same periodicity of the
simulation unit cell, thus it cannot be obtained as a solution of
a Poisson equation that obeys periodic boundary conditions; second,
it is usually computationally convenient to exploit the Fourier-transform
approach of perfectly periodic systems as derived in Eq. \eqref{eq:potential_pbc_full},
thus automatically introducing spurious interactions with periodic
replicas of the unit cell. 

The two shortcomings discussed above can be solved independently.
In particular, auxiliary-function methods are able to screen in reciprocal
space the long range part of the electrostatic potential. Thus interactions
with spurious periodic replicas are removed, even though the computed
potential still retains the (incorrect) periodicity of the simulation
cell. On the other hand, since the system is anyway confined in a
restricted part of the simulation cell, in order to have a correct
estimate of the electrostatic energy it is not necessary to have the
electrostatic potential described accurately everywhere in the unit
cell, but it is only important to have the correct potential in the
region where the source charges are located. For this reason, the
isolated system of interest is usually treated inside large super-cells,
in such a way that deformations of the potential due to the boundary
of the cell do not affect the calculation of the electrostatic energy
of the system. We note in passing that an alternative real-space approach
has been recently proposed that is able to recover the ideal potential
of the system in a computationally effective way, by using a multi-grid
method to correct the 3D FFT-based potential \cite{dabo_prb_2008,dabo_prb_2008_erratum}.

\subsection{Electrostatics in dielectric environments and periodic boundary conditions}

We summarize here the main equations behind continuum solvation, and
in particular as embodied in the SCCS model \cite{andreussi_jcp_2012}.
The quantum-mechanical system of interest is immersed in a dielectric
medium characterized by a density-dependent dielectric constant. A
dielectric function is defined in order to ensure that the dielectric
constant is equal to one in the interior of the solute, where the
electronic density is high, and smoothly acquires the value of the
bulk dielectric permittivity of the solvent, $\epsilon_{0}$, where
the electronic density goes to zero. An optimal definition of the
dielectric function was provided in Ref. \cite{andreussi_jcp_2012}
in terms of only two tunable thresholds. For the sake of simplicity,
in our notation in the following we will not highlight the specific
functional definition of the dielectric function $\epsilon\left[\rho^{el}\left(\mathbf{r}\right)\right]$,
and only consider it as a continuous function, $\epsilon\left(\mathbf{r}\right)$,
defined everywhere in the simulation cell. 

In the presence of a dielectric continuum, the electrostatic potential
will be the solution of the generalized Poisson equation 
\begin{equation}
\nabla\cdot\epsilon\left(\mathbf{r}\right)\nabla v^{\epsilon}\left[\rho^{solute}\right]\left(\mathbf{r}\right)=-4\pi\rho^{solute}\left(\mathbf{r}\right),\label{eq:poisson_dielectric}
\end{equation}
where the superscript $\epsilon$ has been added to distinguish the
potential from the one computed in vacuum. By introducing a polarization
charge density 
\begin{multline}
\rho^{pol}\left(\mathbf{r}\right)=\mathbf{\nabla}\cdot\left(\frac{\epsilon\left(\mathbf{r}\right)-1}{4\pi}\mathbf{\nabla}v^{\epsilon}\left[\rho^{solute}\right]\left(\mathbf{r}\right)\right)\\
=\frac{1}{4\pi}\mathbf{\nabla}\ln\epsilon\left(\mathbf{r}\right)\cdot\mathbf{\nabla}v^{\epsilon}\left[\rho^{solute}\right]\left(\mathbf{r}\right)-\frac{\epsilon\left(\mathbf{r}\right)-1}{\epsilon\left(\mathbf{r}\right)}\rho^{solute}\left(\mathbf{r}\right).\label{eq:rhopol_isol}
\end{multline}
the generalized Poisson equation in a dielectric can be recast into
a vacuum-like Poisson equation
\begin{multline}
\nabla^{2}v^{\epsilon}\left[\rho^{solute}\right]\left(\mathbf{r}\right)=-4\pi\left(\rho^{solute}\left(\mathbf{r}\right)+\rho^{pol}\left(\mathbf{r}\right)\right)\\
=-4\pi\rho^{tot}\left(\mathbf{r}\right),\label{eq:poisson_dielectric_pol}
\end{multline}
that depends self-consistently on the polarization charge density
(and thus on $v^{\epsilon}$ itself), where the electrostatic potential
$v^{\epsilon}$ can be expressed as a vacuum potential depending on
both the source and polarization charge densities
\begin{multline}
v^{\epsilon}\left[\rho^{solute}\right]\left(\mathbf{r}\right)=v\left[\rho^{solute}+\rho^{pol}\right]\left(\mathbf{r}\right)\\
=v\left[\rho^{tot}\right]\left(\mathbf{r}\right)=v\left[\rho^{solute}\right]\left(\mathbf{r}\right)+v\left[\rho^{pol}\right]\left(\mathbf{r}\right).
\end{multline}

From the knowledge of the electrostatic field, one can derive in a
straightforward way the Kohn-Sham potential, the electrostatic energy
and the forces acting on the nuclei, as shown in Ref.\cite{andreussi_jcp_2012}.
In particular, the total electrostatic energy of the system can be
separated into two contributions 
\begin{multline}
E^{\epsilon}\left[\rho^{solute}\right]=\frac{1}{2}\int_{D}\rho^{solute}\left(\mathbf{r}\right)v^{\epsilon}\left[\rho^{solute}\right]\left(\mathbf{r}\right)\mbox{d}\mathbf{r}\\
=\frac{1}{2}\int_{D}\rho^{solute}\left(\mathbf{r}\right)v\left[\rho^{solute}\right]\left(\mathbf{r}\right)\mbox{d}\mathbf{r}\\
+\frac{1}{2}\int_{D}\rho^{solute}\left(\mathbf{r}\right)v\left[\rho^{pol}\right]\left(\mathbf{r}\right)\mbox{d}\mathbf{r}\\
=E\left[\rho^{solute}\right]+E^{pol}\left[\rho^{solute},\rho^{pol}\right],\label{eq:energy_diel}
\end{multline}
where we decided to indicate explicitly the dependence of the second
contribution on the polarization charge density
\begin{multline}
E^{pol}\left[\rho^{solute},\rho^{pol}\right]=\frac{1}{2}\int_{D}\rho^{solute}\left(\mathbf{r}\right)v\left[\rho^{pol}\right]\left(\mathbf{r}\right)\mbox{d}\mathbf{r}\\
=\frac{1}{2}\int_{D}\rho^{pol}\left(\mathbf{r}\right)v\left[\rho^{solute}\right]\left(\mathbf{r}\right)\mbox{d}\mathbf{r}.\label{eq:energy_pol_isol}
\end{multline}

For isolated systems, the Poisson equation should be solved together
with boundary conditions of vanishing potential at long distances.
Nonetheless, most of the approaches proposed in the literature in
order to solve Eq. \eqref{eq:poisson_dielectric} or Eq. \eqref{eq:poisson_dielectric_pol}
introduce some approximations on the boundary conditions, in order
to simplify or speed up the calculation. In particular, in the original
formulation of Fattebert and Gygi \cite{fattebert_jcomputchem_2002,fattebert_intjqchem_2003}
and in some of its following implementations \cite{scherlis_jcp_2006,sanchez_jcp_2009,dziedzic_epl_2011},
a multi-grid method was used to solve for the electrostatic potential,
together with an arbitrary homogeneous zeroing of the potential at
the boundary of the simulation cell (Dirichelet boundary conditions).
In the recently developed SCCS method, instead, an iterative approach
has been proposed, coupled with standard FFTs and which relies on
periodic boundary conditions. 

In particular, one can approximate the isolated potential $v\left[\rho^{tot}\right]\left(\mathbf{r}\right)$
by the periodic potential $\mathtt{v}\left[\varrho^{tot}\right]\left(\mathbf{r}\right)$,
which can be computed in reciprocal space by exploiting Eq. \eqref{eq:potential_pbc}
as
\begin{equation}
\mathtt{v}\left[\varrho^{tot}\right]\left(\mathbf{r}\right)=\sum_{\mathbf{g}\neq0}\frac{4\pi}{g^{2}}\tilde{\varrho}^{tot}\left(\mathbf{g}\right)e^{i\mathbf{g\cdot r}}.\label{eq:potential_pbc_pol}
\end{equation}
where the total charge density $\varrho^{tot}\left(\mathbf{r}\right)$
is also different from the ideal isolated one $\rho^{tot}\left(\mathbf{r}\right)$,
due to its periodicity and of being optimized in the presence of periodic
boundary conditions. While the effect of periodicity on the optimization
of the nuclear and ionic degrees of freedom of a system can be considered
to be negligible \cite{makov_prb_1995}, periodic boundary conditions
enter directly in the definition of the polarization charge density,
due to its dependence on the gradient of the electrostatic field 
\begin{equation}
\mathbf{\nabla}\mathtt{v}\left[\varrho^{tot}\right]\left(\mathbf{r}\right)=\sum_{\mathbf{g}}\frac{4\pi i\mathbf{g}}{g^{2}}\tilde{\varrho}^{tot}\left(\mathbf{g}\right)e^{i\mathbf{g\cdot r}}.
\end{equation}
Moreover, when charged solutes are treated, i.e. when
\begin{equation}
\int_{\mathtt{D}}\varrho^{solute}\left(\mathbf{r}\right)\mathrm{d}\mathbf{r}=\mathtt{q}^{solute}\neq0,
\end{equation}
the presence of the compensating NCB background should be explicitly
accounted for in using the approximation in Eq. \eqref{eq:potential_pbc_pol}.
The polarization charge in the most general case of a charged system
in its periodic approximation is thus given by
\begin{multline}
\varrho^{pol}\left(\mathbf{r}\right)=\frac{1}{4\pi}\mathbf{\nabla}\ln\epsilon\left(\mathbf{r}\right)\cdot\mathbf{\nabla}\mathtt{v}\left(\varrho^{tot}-\left\langle \varrho^{tot}\right\rangle ,\mathbf{r}\right)\\
-\frac{\epsilon\left(\mathbf{r}\right)-1}{\epsilon\left(\mathbf{r}\right)}\varrho^{solute}\left(\mathbf{r}\right)+\frac{\epsilon\left(\mathbf{r}\right)-1}{\epsilon\left(\mathbf{r}\right)}\left\langle \varrho^{solute}\right\rangle .\label{eq:rhopol_pbc}
\end{multline}

Similarly to the case of a polarization charge density in vacuum,
the first two terms of $\varrho^{pol}$ are localized in the narrow
transition region at the boundary of the solute, as explained in Ref
\cite{andreussi_jcp_2012}. On the contrary, the last contribution
appearing in Eq. \eqref{eq:rhopol_pbc} is defined everywhere in the
simulation cell, except for the vacuum region inside the solute, where
$\epsilon\left(\mathbf{r}\right)=1$. 

It is important to notice that, even though for an isolated charged
solute Gauss's law would require the total polarization charge to
fulfill the following sum rule
\begin{equation}
\int_{D}\rho^{pol}\left(\mathbf{r}\right)d\mathbf{r}=-\frac{\epsilon_{0}-1}{\epsilon_{0}}q^{solute},
\end{equation}
the total polarization charge of a system in periodic boundary conditions
will sum up to zero
\begin{equation}
\int_{D}\varrho^{pol}\left(\mathbf{r}\right)d\mathbf{r}=0,
\end{equation}
due to the PBC-imposed neutrality of the source charge density. 

Provided that the full Eq. \eqref{eq:rhopol_pbc} is used to compute
the polarization density, all the equations derived in Ref. \cite{andreussi_jcp_2012}
apply straightforwardly. For neutral solutes immersed in solvents
with high dielectric permittivity and reasonably large cell sizes,
the effect of PBC was already shown to be negligible (see Figure (17)
of Ref.\cite{andreussi_jcp_2012}). Nonetheless, charged systems immersed
in solvents with low dielectric permittivitys may present a substantial
dependence on the size of the simulation cells.

\subsection{Makov-Payne correction in dielectric environments}

To summarize the previous discussion, when approximating an isolated
system with its periodic counterpart in a quantum-mechanical calculation,
one is actually performing two different approximations. 
\begin{itemize}
\item First, 
\begin{equation}
\varrho\left(\mathbf{r}\right)\neq\rho\left(\mathbf{r}\right)
\end{equation}
i.e. the charge density that one is optimizing with PBC will in general
converge to a different final state from the ideal isolated case,
due to the interaction with the periodic images and the neutralizing
charge backgound (NCB). 
\item Second, 
\begin{equation}
\mathtt{v}\left[\rho\right]\left(\mathbf{r}\right)\neq v\left[\rho\right]\left(\mathbf{r}\right)
\end{equation}
i.e. even assuming that we are dealing with a neutral system and that
the effects of periodicity on its optimized charge density are negligible,
the periodic potential will be different from the isolated case due
to the contributions arising from the periodic images and, possibly,
due to the different boundary conditions used to solve the problem. 
\end{itemize}
Both approximations will contribute to an error in the calculation
of the total energy, i.e. 
\begin{equation}
\mathtt{E}\left[\varrho\right]\neq\mathtt{E}\left[\rho\right]\neq E\left[\rho\right].
\end{equation}
Nonetheless, a simple analytical expression for the leading contributions
to the difference between the above energies can be derived in the
special case of a cubic simulation cell. The first derivation of such
an expression is due to Makov and Payne \cite{makov_prb_1995} and
provides an approximation of $E\left[\rho\right]$ whose system size
dependence is at worst of the order of $L^{-5}$, where $L$ is the
size of the cubic cell. Namely, the exact electrostatic energy of
the isolated system can be written in terms of its periodic approximation
as 
\begin{multline}
E^{solute}\left[\rho^{solute}\right]=\\
=\mathtt{E}^{\mathtt{solute}}\left[\varrho^{\mathtt{solute}}-\left\langle \varrho^{\mathtt{solute}}\right\rangle \right]+\frac{\left(q^{solute}\right)^{2}\alpha_{0}}{2L}\\
-\frac{2\pi}{3L^{3}}\left(q^{solute}Q^{solute}-\left(d^{solute}\right)^{2}\right)+O\left(L^{-5}\right),\label{eq:makov-payne_vacuum}
\end{multline}
where, with respect to Eq. (15) of Ref. \cite{makov_prb_1995}, the
second contribution has the correct sign and is expressed explicitly
in terms of the isolated solute multipole moments. Moreover, the Makov-Payne
derivation correctly assumes that charge relaxation due to the artificial
periodicity of the system only contributes to the correction of the
energy at higher orders. Thus, the multipole moments that enter Eq.
\eqref{eq:makov-payne_vacuum} are computed from the periodic density
in the unit cell without including the eventual NCB density 
\begin{align}
q^{solute} & \approx\int_{\mathtt{D}}\varrho^{\mathtt{solute}}\left(\mathbf{r}\right)\mbox{d}\mathbf{r}\equiv\mathtt{q^{solute-ncb}}\label{eq:solute_charge_relax}\\
\mathbf{d}^{solute} & \approx\int_{\mathtt{D}}\varrho^{\mathtt{solute}}\left(\mathbf{r}\right)\mathbf{r}\mbox{d}\mathbf{r}\equiv\mathtt{d^{solute-ncb}}\label{eq:solute_dipole_relax}\\
Q^{solute} & \approx\int_{\mathtt{D}}\varrho^{\mathtt{solute}}\left(\mathbf{r}\right)\mathbf{r}^{2}\mbox{d}\mathbf{r}\equiv\mathtt{Q^{solute-ncb}}.\label{eq:solute_quad_relax}
\end{align}
The first contribution in Eq. \eqref{eq:makov-payne_vacuum} is due
to the interaction energy of the NCB-neutralized monopole moment in
the periodic system interacting with its replicas and is easily expressed
in terms of the Madelung constant of a cubic lattice $\alpha_{0}$.
Dipole-dipole and quadrupole-monopole interactions with periodic replicas
are canceled in the periodic energy due to the cubic symmetry of the
lattice, while the contributions due to quadrupole-quadrupole and
higher multipoles interactions decay at worst as $L^{-5}$. The origin
of the second contribution in Eq. \eqref{eq:makov-payne_vacuum} is
due to the tin-foil boundary conditions that are implicitly assumed
in a periodic boundary calculation. These boundary conditions artificially
impose that the average electrostatic field and potential in the cell
are zero. As a consequence, the interaction energies of the multipole
moments of the system with themselves (specifically the dipole-dipole
and the monopole-quadrupole interactions) are modified with respect
to the isolated case due to these arbitrary shifts. In particular,
the energy due to the dipole moment 
\begin{equation}
\mathtt{E}\left[\mathbf{d}^{solute}\right]=-\frac{1}{2}\mathbf{d}^{solute}\cdot\boldsymbol{\Xi}\left(0\right)
\end{equation}
lacks the contribution 
\begin{multline}
\Delta E=\mathtt{E}\left[\mathbf{d}^{solute}\right]-E\left[\mathbf{d}^{solute}\right]=\\
-\frac{1}{2}\mathbf{d}^{solute}\cdot\left\langle \mathbf{E}\right\rangle =-\frac{1}{2}\mathbf{d}^{solute}\cdot\frac{4\pi}{3}\frac{\mathbf{d}^{solute}}{L^{3}}.
\end{multline}
Similarly, the energy due to the monopole-quadrupole interaction 
\begin{equation}
E=q^{solute}\mathtt{v}^{Q^{solute}}\left(0\right),
\end{equation}
has to be corrected due to the shift of the potential with respect
to the ideal isolated system in vacuum, namely 
\begin{equation}
E^{mq,corr}=q^{solute}\left\langle \mathtt{v}^{Q^{solute}}\right\rangle =q^{solute}\frac{2\pi}{3}\frac{Q^{solute}}{L^{3}}.
\end{equation}
In the above equations we used the fact that, as shown in Ref. \cite{heinz_jcp_2005}
and reported in Eqs. \eqref{eq:field_ext} and \eqref{eq:potential_ext},
only the dipole and quadrupole moments contribute to the average values
of the electrostatic field and potential, respectively, i.e. 
\begin{align}
\left\langle \mathbf{\boldsymbol{\Xi}}^{\rho^{solute}}\right\rangle  & =\left\langle \mathbf{\boldsymbol{\Xi}}^{\mathbf{d}^{solute}}\right\rangle =\mathbf{\boldsymbol{\Xi}}_{0}\\
\left\langle v^{\rho^{solute}}\right\rangle  & =\left\langle v^{Q^{solute}}\right\rangle =v_{0}.
\end{align}

Makov and Payne also derived a simplified expression for a system
in a condensed phase, by adopting the approach of Leslie and Gillan
and rescaling the potential by the dielectric constant $\epsilon_{0}$
of the system. The result 
\begin{multline}
E^{solute}\left[\rho^{solute}\right]=\mathtt{E}^{\mathtt{solute}}\left[\varrho^{\mathtt{solute}}\right]+\frac{\left(q^{\mathtt{solute}}\right)^{2}\alpha_{0}}{2L\epsilon_{0}}\\
-\frac{2\pi}{3L^{3}\epsilon_{0}}\left(q^{\mathtt{solute}}\mathtt{Q}^{\mathtt{solute}}-\left(\mathtt{d}^{\mathtt{solute}}\right)^{2}\right)+O\left(L^{-5}\right),\label{eq:makov-payne_diel}
\end{multline}
assumes a uniform homogeneous dielectric everywhere in space. Such
an assumption does not take into account the variations of the dielectric
constant in the different regions of the system studied and, in particular,
is not correct for the SCCS model, where a solute is immersed in a
medium whose dielectric constant varies from one (vacuum) to the bulk
dielectric constant of the solvent. Nonetheless, an approach similar
to the one of Makov and Payne can be used to derive the correction
to the electrostatic energy in the SCCS framework up to terms of the
order of $L^{-3}$. 

Contrary to what is generally assumed for the energy contribution
due to the polarization of the solute charge density due to periodic
images, the periodic solution of the polarization charge density has
a significant effect on the polarization contribution to the electrostatic
energy of a solvated system. The problem is twofold, and is partly
related to the fact that the neutralizing background induces a small
polarization which is diffused all over the unit cell, and partly
due to the fact that periodic images can induce a non-negligible polarization
charge density in the region of space close to the solute charge density.
These spurious polarization charges affect the multipole moments of
the polarization charge 
\begin{align}
\mathtt{q}^{\mathtt{pol-j}} & \neq q^{pol}\\
\mathtt{d}^{\mathtt{pol-j}} & \neq\mathbf{d}^{pol}\\
\mathtt{Q}^{\mathtt{pol-j}} & \neq Q^{pol}.
\end{align}
and need to be taken care of explicitly, before a scheme analogous
to the one of Makov and Payne can be adopted. 

In particular, the difference $\Delta\rho\left(\mathbf{r}\right)$
between the periodic (Eq. \eqref{eq:rhopol_pbc}) and the isolated
(Eq. \eqref{eq:rhopol_isol}) polarization charges can be written
as
\begin{multline}
\Delta\rho^{pol}\left(\mathbf{r}\right)\equiv\varrho^{pol}\left(\mathbf{r}\right)-\rho^{pol}\left(\mathbf{r}\right)\\
=\frac{1}{4\pi}\mathbf{\nabla}\ln\epsilon\left(\mathbf{r}\right)\cdot\left[\nabla\mathtt{v}\left[\varrho^{tot}-\left\langle \varrho^{tot}\right\rangle \right]\left(\mathbf{r}\right)\right.\\
\left.-\mathbf{\nabla}v\left[\rho^{tot}\right]\left(\mathbf{r}\right)\right]+\frac{\epsilon\left(\mathbf{r}\right)-1}{\epsilon\left(\mathbf{r}\right)}\left\langle \varrho^{solute}\right\rangle \approx\\
\approx\frac{1}{4\pi}\mathbf{\nabla}\ln\epsilon\left(\mathbf{r}\right)\cdot\mathbf{\nabla}v\left[\Delta\rho^{pol}\right]\left(\mathbf{r}\right)+\frac{\epsilon\left(\mathbf{r}\right)-1}{\epsilon\left(\mathbf{r}\right)}\left\langle \varrho^{solute}\right\rangle \\
+\frac{1}{4\pi}\mathbf{\nabla}\ln\epsilon\left(\mathbf{r}\right)\cdot\mathbf{\nabla}\Delta v\left[\varrho^{tot}\right]\left(\mathbf{r}\right).\label{eq:delta_rhopol}
\end{multline}
The corrective potential $\Delta v\left[\varrho\right]\left(\mathbf{r}\right)$
is defined following Dabo et al. \cite{dabo_prb_2008,dabo_prb_2008_erratum},
i.e. as the difference between the ideal isolated potential in vacuum
and its periodic counterpart computed using tin-foil boundary conditions
\begin{equation}
\Delta v\left[\varrho\right]\left(\mathbf{r}\right)=\mathtt{v}\left[\varrho-\left\langle \varrho\right\rangle \right]\left(\mathbf{r}\right)-v\left[\varrho\right]\left(\mathbf{r}\right)\approx\mathtt{v}\left[\varrho-\left\langle \varrho\right\rangle \right]\left(\mathbf{r}\right)-v\left[\varrho\right]\left(\mathbf{r}\right).
\end{equation}
The correction to the polarization charge can be computed iteratively
with the same approach used for the periodic polarization charge,
provided that an expression for the corrective potential is available.
The last two contributions to the polarization in Eq. \eqref{eq:delta_rhopol}
do not change during the iteration cycles, thus they can be considered
as two separate sources and the corrective polarization can be separated
into two contributions, one due to the NCB density and the other due
to the corrective potential. Namely
\begin{multline}
\Delta\rho^{pol,ncb}\left(\mathbf{r}\right)=\frac{1}{4\pi}\mathbf{\nabla}\ln\epsilon\left(\mathbf{r}\right)\cdot\mathbf{\nabla}v\left[\Delta\rho^{pol,ncb}\right]\left(\mathbf{r}\right)\\
+\frac{\epsilon\left(\mathbf{r}\right)-1}{\epsilon\left(\mathbf{r}\right)}\left\langle \varrho^{solute}\right\rangle 
\end{multline}
and
\begin{multline}
\Delta\rho^{pol,periodic}\left(\mathbf{r}\right)=\frac{1}{4\pi}\mathbf{\nabla}\ln\epsilon\left(\mathbf{r}\right)\cdot\mathbf{\nabla}v\left[\Delta\rho^{pol,periodic}\right]\left(\mathbf{r}\right)\\
+\frac{1}{4\pi}\mathbf{\nabla}\ln\epsilon\left(\mathbf{r}\right)\cdot\mathbf{\nabla}\Delta v\left[\varrho^{tot}\right]\left(\mathbf{r}\right)
\end{multline}
By exploiting the derivation of Ref. \cite{dabo_prb_2008,dabo_prb_2008_erratum}
for the point-charge approximation of the corrective potential (see
following Section), the gradient in the second term of the difference
between periodic and isolated polarization can be approximated as
\begin{equation}
\nabla\Delta v\left[\varrho\right]\left(\mathbf{r}\right)\approx\frac{4\pi}{3L^{3}}\left(\mathbf{d}-\mathbf{r}\right).
\end{equation}
It is important to note that the above approximation is correct only
close to the origin of the system charge distribution and it becomes
more and more accurate as the cell size increases. Both the periodic
and the NCB contribution to the corrective polarization charge are
proportional to $L^{-3}$. While the periodic polarization is defined
only in the small region around the solute, where the dielectric is
allowed to vary, the NCB polarization is defined everywhere in space.
Nonetheless, its value in the bulk of the solvent is constant and
given by 
\begin{equation}
\Delta\rho^{pol,ncb,bulk}=\frac{\left(\epsilon_{0}-1\right)}{\epsilon_{0}}\left\langle \varrho^{solute}\right\rangle .
\end{equation}
The bulk constant charge can be removed from the corrective polarization
so that 
\begin{multline}
\Delta\rho^{pol,ncb,confined}=\frac{1}{4\pi}\mathbf{\nabla}\ln\epsilon\left(\mathbf{r}\right)\cdot\mathbf{\nabla}v\left[\Delta\rho^{pol,ncb}\right]\left(\mathbf{r}\right)\\
+\left(\frac{1}{\epsilon_{0}}-\frac{1}{\epsilon\left(\mathbf{r}\right)}\right)\frac{q^{solute}}{L^{3}}
\end{multline}
is a quantity confined in a well defined region of space, which do
not depend on cell size. With this choice, the energy contributions
due to the corrective polarizations are 
\begin{multline}
\int_{\mathtt{D}}\Delta\rho^{pol}\left(\mathbf{r}\right)\mathtt{v}\left[\rho^{solute}\right]\left(\mathbf{r}\right)\mbox{d}\mathbf{r}=\\
\int_{\mathtt{D}}\Delta\rho^{pol,periodic}\left(\mathbf{r}\right)\mathtt{v}\left[\rho^{solute}\right]\left(\mathbf{r}\right)\mbox{d}\mathbf{r}\\
+\int_{\mathtt{D}}\Delta\rho^{pol,ncb,confined}\left(\mathbf{r}\right)\mathtt{v}\left[\rho^{solute}\right]\left(\mathbf{r}\right)\mbox{d}\mathbf{r}\\
+\int_{\mathtt{D}}\Delta\rho^{pol,ncb,bulk}\left(\mathbf{r}\right)\mathtt{v}\left[\rho^{solute}\right]\left(\mathbf{r}\right)\mbox{d}\mathbf{r}=\\
=\int_{\mathtt{D}}\Delta\rho^{pol,periodic}\left(\mathbf{r}\right)\mathtt{v}\left[\rho^{solute}\right]\left(\mathbf{r}\right)\mbox{d}\mathbf{r}\\
+\int_{\mathtt{D}}\Delta\rho^{pol,ncb,confined}\left(\mathbf{r}\right)\mathtt{v}\left[\rho^{solute}\right]\left(\mathbf{r}\right)\mbox{d}\mathbf{r}\label{eq:energy_delta_rhopol}
\end{multline}
where the NCB-bulk contribution vanished, as a constant charge density
does not contribute to the periodic energy. Both corrective contributions
in Eq. \eqref{eq:energy_delta_rhopol} will scale as $L^{-3}$, since
the charge densities are confined in a region of space which is not
dependent on the cell size. Thus, when trying to remove the system's
size dependence from the calculation, both terms should be subtracted
from the periodic polarization energy computed in the SCCS framework
\begin{equation}
\mathtt{E}\left[\varrho^{\mathtt{solute}},\rho^{pol}\right]=\mathtt{E}\left[\varrho^{\mathtt{solute}},\varrho^{\mathtt{pol}}\right]-\mathtt{E}^{\mathtt{pol}}\left[\varrho^{\mathtt{solute}},\Delta\rho^{pol}\right].
\end{equation}

Eventually, we are left with the periodic energy of the solute in
vacuum $\mathtt{E^{solute}}\left[\varrho^{\mathtt{solute}}\right]$
(whose ideally isolated counterpart $E^{solute}\left[\rho^{solute}\right]$
can be recovered through the Makov-Payne expression) plus the periodic
energy of interaction between the solute charge density and a polarization
optimized as if the system were isolated
\begin{multline}
\mathtt{E}^{pol}\left[\varrho^{solute},\rho^{pol}\right]=\frac{1}{2}\int\rho^{pol}\left(\mathbf{r}\right)\mathtt{v}\left[\varrho^{solute}\right]\left(\mathbf{r}\right)\mbox{d}\mathbf{r}\\
=\frac{1}{2}\int\varrho^{solute}\left(\mathbf{r}\right)\mathtt{v}\left[\rho^{pol}\right]\left(\mathbf{r}\right)\mbox{d}\mathbf{r}.\label{eq:energy_pol}
\end{multline}

A Makov-Payne like expression for this latter term can be derived
by assuming, as in Ref. \cite{makov_prb_1995}, that $\varrho^{solute}\left(\mathbf{r}\right)=\rho^{solute}\left(\mathbf{r}\right)$
inside the unit cell and by considering the Makov-Payne corrections
for the electrostatic energy of the system composed by the total charge
density 
\begin{equation}
\rho^{tot}\left(\mathbf{r}\right)=\rho^{sol}\left(\mathbf{r}\right)+\rho^{pol}\left(\mathbf{r}\right)
\end{equation}

and the one of a system composed solely by the polarization charge.
Namely, by rewriting Eq. \eqref{eq:makov-payne_vacuum} for the total
charge distribution and by performing some simple algebraic manipulation
one obtains 
\begin{multline}
E^{pol}\left[\rho^{solute},\rho^{pol}\right]=\\
=\frac{1}{2}\left(E^{solute}\left[\rho^{tot}\right]-E^{solute}\left[\rho^{solute}\right]-E^{solute}\left[\rho^{pol}\right]\right)\\
=\mathtt{E}^{\mathtt{pol}}\left[\varrho^{\mathtt{solute}},\rho^{pol}\right]+\frac{\left(q^{solute}\right)^{2}\alpha_{0}}{2L}\left(-1+\frac{1}{\epsilon_{0}}\right)\\
-\frac{\pi q^{solute}}{3L^{3}}\left(-\frac{\epsilon_{0}-1}{\epsilon_{0}}\mathtt{Q}^{\mathtt{solute}}+Q^{pol}\right)\\
+\frac{2\pi}{3L^{3}}\left(\mathbf{d}^{\mathtt{solute}}\cdot\mathbf{d}^{pol}\right)
\end{multline}
where the relation 
\begin{equation}
q^{pol}=-\frac{\epsilon_{0}-1}{\epsilon_{0}}q^{solute}
\end{equation}
has been exploited between the total polarization charge in the isolated
system and the solute charge. When summing the correction to the polarization
energy to the one of the electrostatic energy of the system in vacuum,
Eq. \eqref{eq:makov-payne_vacuum}, the final expression for the energy
of the solvated system becomes
\begin{multline}
E\left[\rho^{solute},\rho^{pol}\right]=\mathtt{E}\left[\varrho^{solute},\varrho^{pol}\right]-\mathtt{E}^{\mathtt{pol}}\left[\varrho^{\mathtt{solute}},\Delta\rho^{pol}\right]\\
+\left(\frac{1}{\epsilon_{0}}\right)\frac{\left(q^{solute}\right)^{2}\alpha_{0}}{2L}\\
-\frac{2\pi q^{solute}}{3L^{3}}\left(\frac{\mathtt{Q}^{\mathtt{solute}}}{2\epsilon_{0}}+\frac{\mathtt{Q}^{\mathtt{solute}}+Q^{pol}}{2}\right)\\
+\frac{2\pi}{3L^{3}}\left(\left(\mathtt{d}^{solute}\right)^{2}+\mathbf{d}^{solute}\cdot\mathbf{d}^{pol}\right).\label{eq:makov-payne_sccs}
\end{multline}

Compared to the result derived by Makov and Payne for aperiodic systems
in a condensed phase, we see that the monopole contribution is identical,
reflecting the fact that it is an interaction energy between systems
in neighboring cells and is thus exactly rescaled by the presence
of the dielectric continuum. On the other hand a more complex expression
for the other terms has now been obtained in Eq. \ref{eq:makov-payne_sccs},
and this is one of the main results of this paper. 

If we consider the simple case of a uniform dielectric extending over
the whole space, the polarization charge would be simply given by
\begin{equation}
\rho^{pol}\left(\mathbf{r}\right)=-\frac{\epsilon_{0}-1}{\epsilon_{0}}\rho^{solute}\left(\mathbf{r}\right),
\end{equation}
which translates into dipole and quadrupole moments 
\begin{align}
\mathbf{d}^{pol} & =-\frac{\epsilon_{0}-1}{\epsilon_{0}}\mathbf{d}^{solute}\\
Q^{pol} & =-\frac{\epsilon_{0}-1}{\epsilon_{0}}Q^{solute},
\end{align}
which, inserted in Eq. \eqref{eq:makov-payne_sccs}, give back the
result proposed by Makov and Payne, Eq \eqref{eq:makov-payne_diel}.
In the case of a dipolar solute in a spherical cavity, the polarization
dipole is analytically obtained from the Onsager model as 
\begin{equation}
\mathbf{d}^{pol,Onsager}=-\frac{2\left(\epsilon_{0}-1\right)}{2\epsilon_{0}+1}\mathbf{d}^{solute},
\end{equation}
which gives a term proportional to
\begin{equation}
\frac{2\pi}{\left(2\epsilon_{0}+1\right)L^{3}}\left(d^{solute}\right)^{2},
\end{equation}
consistent with the expression of the extrinsic field of a periodic
system immersed in a dielectric, $\mathbf{E}_{0}^{\epsilon}$ in Eq.
\eqref{eq:field_ext_diel}. In general, for arbitrary, molecular shaped
cavities, the dipole and quadrupole contributions are not analytic
functions of the solute multipole moments and need to be computed
explicitly from the integral of the polarization charge density.

\subsection{Counter-charge corrections in dielectric environments}

Several schemes have been proposed along the lines of the Makov-Payne
correction, but which self-consistently correct the electrostatic
potential rather than just the final electrostatic energy. The general
framework is to recover the electrostatic potential of the isolated
system by adding to the periodic boundary potential a corrective term.
The correction can then be analytically computed for specific approximations
on the charge distribution of the system, or the exact problem can
be solved numerically via multi-grid techniques. The different schemes
have been recently classified into three categories, depending on
the different level of approximations used to tread the charge density
of the system: following Ref. \cite{dabo_prb_2008,dabo_prb_2008_erratum},
they are labelled as point countercharge (PCC), Gaussian countercharge
(GCC), and density countercharge (DCC) methods. Here we will discuss
the PCC correction scheme. For a point-like unit charge in a cubic
cell, the corrective potential 
\begin{equation}
\Delta v\left[\rho\right]\left(\mathbf{r}\right)=\frac{\alpha_{0}}{L}-\frac{2\pi}{3L^{3}}r^{2}+O\left(\left|\mathbf{r}^{4}\right|\right)
\end{equation}
can be recovered by exploiting symmetry and the Poisson Eq. \eqref{eq:poisson_vacuum},
as shown in Ref. \cite{dabo_prb_2008,dabo_prb_2008_erratum}. The
resulting parabolic potential is accurate only close to where the
charge is located. For an arbitrary charge distribution, the corrective
potential can be expressed in terms of the corrective potential of
a collection of point-charges that matches the system's multipole
moments. Due to the quadratic nature of the correction, only the potential
generated by multipoles up to the quadrupole can be corrected. The
final PCC expression for the corrective potential reads 
\begin{multline}
\Delta v\left[\rho\right]\left(\mathbf{r}\right)=\frac{\alpha_{0}}{L}q-\frac{2\pi q}{3L^{3}}r^{2}+\frac{4\pi}{3L^{3}}\mathbf{d}\cdot\mathbf{r}-\frac{2\pi}{3L^{3}}Q.\label{eq:deltav_pcc}
\end{multline}
The correction to the energy 
\begin{multline}
\Delta E=E^{pol}\left[\rho\right]-\mathtt{E}\left[\rho\right]\\
=\frac{1}{2}\int_{\mathtt{D}}\rho\Delta v\left[\rho\right]\left(\mathbf{r}\right)\\
=\frac{\alpha_{0}}{2L}q^{2}-\frac{2\pi}{3L^{3}}\left(qQ-d^{2}\right)
\end{multline}
reduces correctly to the Makov-Payne expression, with the only difference
that the molecular charge distribution is now optimized in the presence
of the corrected potential, i.e. the approximations in Eqs. \eqref{eq:solute_charge_relax},
\eqref{eq:solute_dipole_relax}, and \eqref{eq:solute_quad_relax}
are not needed. 

When a continuum dielectric is present in the system, the electrostatic
energy is that of Eq. \eqref{eq:energy_diel}, and the potential that
needs to be corrected is the one arising from the total charge distribution,
\begin{equation}
\mathtt{v}^{\epsilon}\left[\rho^{solute}\right]\left(\mathbf{r}\right)=\mathtt{v}\left[\rho^{solute}+\rho^{pol}\right]\left(\mathbf{r}\right)=\mathtt{v}\left[\rho^{tot}\right]\left(\mathbf{r}\right),
\end{equation}
 including the polarization charge. This means that Eq.\ref{eq:deltav_pcc}
can be simply modified as 
\begin{multline}
\Delta v\left[\rho^{tot}\right]\left(\mathbf{r}\right)=\frac{\alpha_{0}}{L}q^{tot}-\frac{2\pi q^{tot}}{3L^{3}}r^{2}\\
+\frac{4\pi}{3L^{3}}\mathbf{d}^{tot}\cdot\mathbf{r}-\frac{2\pi}{3L^{3}}Q^{tot}\label{eq:deltav_pcc+sccs_0d}
\end{multline}
where 
\begin{align}
q^{tot} & =q^{sol}+q^{pol}=\frac{q^{sol}}{\epsilon_{0}}\\
\mathbf{d}^{tot} & =\mathbf{d}^{sol}+\mathbf{d}^{pol}\\
Q^{tot} & =Q^{sol}+Q^{pol}.
\end{align}
Again, when computing the correction to the electrostatic energy of
the system, the PCC approach gives the same result obtained with the
Makov-Payne scheme,
\begin{multline}
\Delta E=\frac{1}{2}\int\rho^{solute}\Delta v\left[\rho^{tot}\right]\left(\mathbf{r}\right)\mbox{d}\mathbf{r}=\\
=\frac{\alpha_{0}}{2L}\frac{q^{solute}}{\epsilon_{0}}-\frac{2\pi}{3L^{3}}\left[q^{solute}\left(\frac{Q^{tot}}{2}+\frac{Q^{solute}}{2\epsilon_{0}}\right)+\right.\\
\left.-\left(\mathbf{d}^{tot}\right)\cdot\mathbf{d}^{solute}\right]
\end{multline}
 apart from the fact that the solute charge density is now optimized
in the presence of the correct boundary conditions. From the above
expressions, it is straightforward to derive the correction to the
interatomic forces, namely
\begin{multline}
\Delta\mathbf{f}_{a}=-\frac{d\Delta E}{d\mathbf{R}_{a}}=z_{a}\left.\nabla\Delta v\left[\rho^{tot}\right]\left(\mathbf{r}\right)\right|_{\mathbf{r=R}_{a}}=\\
=z_{a}\frac{4\pi}{3L^{3}}\left(-q^{tot}\mathbf{R}_{a}+\mathbf{d}^{tot}\right)
\end{multline}
where we have used the Hellmann-Feynman theorem, following the derivation
reported in Section IIIC of Ref. \cite{andreussi_jcp_2012}, and for
the solute charge density we have used 
\begin{equation}
\rho^{solute}=\rho^{elec}+\sum_{a}z_{a}\delta\left(\mathbf{r}-\mathbf{R}_{a}\right),
\end{equation}
where the nuclei are represented as point-like charges.

As we are now using the correct potential in the derivation of the
polarization charges, no NCB polarization and no periodic polarization
appear in the polarization charge. Similarly, provided that the potential
is correct up to the region where the dielectric medium becomes uniform,
the total polarization charge will sum up to the correct value for
an isolated system. As summarized in Appendix A, special care needs
to be taken in the way nuclear charges are treated when computing
the polarization charge and PCC periodic-boundary corrections. 

A similar approach can be derived for systems of different periodicity,
where in particular the exact expression of the corrective potential
can be obtained analytically via partial Fourier transforms. A particularly
important case is the one of two-dimensional systems, for which a
solution involving two-dimensional Fourier transform was derived in
Ref.s \cite{minary_jcp_2002} and \cite{li_prb_2011}. Analogously
to what was done with PCC, a simple approximated analytical solution
can be devised for the case where the cell size is large enough compared
to the size of the system. In this case, only the component for $\mathbf{g}=0$
contributes significantly to the corrective potential, that acquires
a quadratic form analogous to the PCC results reported for the isolated
system in cubic cells. Namely, the expression for the corrective potential
of a 2D system is 
\begin{multline}
\Delta v^{2D}\left[\rho^{tot}\right]\left(\mathbf{r}\right)=\Delta v^{2D}\left[\rho^{tot}\right]\left(r_{z}\right)\\
=\frac{\alpha_{1D}}{L_{z}}q^{tot}-\frac{2\pi q^{tot}}{AL_{z}}r_{z}^{2}+\frac{4\pi}{AL_{z}}d_{z}^{tot}\cdot r_{z}-\frac{2\pi}{AL_{z}}Q_{zz}^{tot},\label{eq:deltav_pcc+sccs_2d}
\end{multline}
where $\alpha_{1D}=\pi/3$ is the Madelung constant of a one-dimensional
periodic array of charges, $A$ is the area of the slab, while $L_{z}$
is the size of the cell axis perpendicular to the plane of the slab.
The correction to the energy is readily obtained by integration with
the system charge density; namely, for a system in vacuum
\begin{equation}
\Delta E^{2D}=\frac{\alpha_{1D}}{2L_{z}}\left(q^{solute}\right)^{2}-\frac{2\pi}{AL_{z}}\left(q^{solute}Q_{zz}^{solute}-\left(d_{z}^{solute}\right)^{2}\right).
\end{equation}

Similarly to what was derived for isolated systems, also for slabs
the effects of the solvent can be immediately included by defining
the corrective potentials in terms of the total dipole moment of the
system, thus including the contribution of the polarization density
\begin{multline}
\Delta E^{2D}=\frac{\alpha_{1D}}{2L_{z}}\frac{\left(q^{solute}\right)}{\epsilon_{0}}^{2}\\
-\frac{2\pi}{AL_{z}}\left(q^{solute}\left(\frac{Q_{zz}^{solute}+Q_{zz}^{pol}}{2}+\frac{Q_{zz}^{solute}}{2\epsilon_{0}}\right)+\right.\\
\left.-\left(d_{z}^{solute}+d_{z}^{pol}\right)d_{z}^{solute}\right).
\end{multline}
For the periodic-boundary correction contribution to inter-atomic
forces an expression similar to the one derived for the 0D case applies,
namely
\begin{multline}
\Delta f_{a,z}=-\frac{d\Delta E^{2D}}{dR_{a,z}}=z_{a}\left.\frac{d}{dr_{z}}\Delta v\left[\rho^{tot}\right]\left(r_{z}\right)\right|_{r_{z}=R_{a,z}}\\
=z_{a}\frac{4\pi}{AL_{z}}\left(-qR_{a,z}+d_{z}\right)
\end{multline}

\subsection{Martyna-Tuckerman corrections in a dielectric environment}

While the approaches derived above aim at correcting the periodic
potential by introducing a real-space potential computed a posteriori,
a different approach has been developed in the literature, which aims
to correct directly the periodic potential as computed in reciprocal
space. Such an approach, which has its foundation in the screening
function formalisms and was pioneered for PBC-correction by Martyna
and Tuckerman \cite{martyna_jcp_1999,minary_jcp_2002,minary_jcp_2004},
has received a lot of attention in recent years due to its simple
implementation and very reduced computational cost. 

The main idea behind the approach of Martyna and Tuckerman (MT) \cite{martyna_jcp_1999}
and similar approaches \cite{li_prb_2011} is the following: when
the electrostatic problem is solved in reciprocal space, the use of
the Fourier transform of the differential operator (Eq. \eqref{eq:ft_gradient})
implies that the potential is obtained from the periodic sum of the
real space potentials. In other words, the analytic Fourier transform
of the aperiodic Green's function $G\left(\mathbf{r}-\mathbf{r}'\right)$
corresponds to the reciprocal space coefficients of the periodic Green's
function $\mathtt{G}\left(\mathbf{g}\right)$, namely 
\begin{multline}
\tilde{G}\left(\mathbf{g}\right)=\int_{\infty}G\left(\mathbf{r}\right)e^{-i\mathbf{g}\cdot\mathbf{r}}\mbox{d}\mathbf{r}=\int_{\mathtt{D}}\sum_{\mathbf{R}}G\left(\mathbf{r+R}\right)e^{-i\mathbf{g}\cdot\mathbf{r}}\mbox{d}\mathbf{r}\\
=\int_{\mathtt{D}}\mathtt{G}\left(\mathbf{r}\right)e^{-i\mathbf{g}\cdot\mathbf{r}}\mbox{d}\mathbf{r}=\mathtt{G}\left(\mathbf{g}\right)
\end{multline}
In particular, for the case of an isolated system in vacuum (Eq. \eqref{eq:green_isol_vacuum})
we have 
\begin{equation}
\tilde{G}\left(\mathbf{g}\right)=\mathtt{G}\left(\mathbf{g}\right)=\frac{1}{g^{2}}.
\end{equation}
If one, instead, were to use the Fourier series coefficients of the
potential kernel,
\begin{equation}
G\left(\mathbf{g}\right)=\int_{\mathtt{D}}\frac{e^{-i\mathbf{g}\cdot\mathbf{r}}}{\left|\mathbf{r}\right|}\mbox{d}\mathbf{r}
\end{equation}
one could build the first-image form of the potential, 
\begin{equation}
\hat{G}\left(\mathbf{r}\right)=\sum_{\mathbf{g}}G\left(\mathbf{g}\right)e^{-i\mathbf{g}\cdot\mathbf{r}}
\end{equation}
i.e. a periodically repeated approximation of the isolated Green's
function $G\left(\mathbf{r}\right)$. The periodicity which is introduced
by using the Fourier series and the first-image form only affects
the potential at the boundary of the simulation cell. For this reason,
in short-ranged functions which decay well within the boundaries of
the unit cell, the first-image form, the true potential and the periodic
sum are identical in the region of interest, i.e.
\begin{equation}
\hat{G}^{short}\left(\mathbf{r}\right)=\mathtt{G}^{short}\left(\mathbf{r}\right)=G^{short}\left(\mathbf{r}\right)\quad\mbox{for }\mathbf{r}\in\mathtt{D}
\end{equation}
For long-ranged functions, as is the case for the Coulomb potential,
it is generally accepted that if a cell twice as large as the system
studied is used, the first-image form is a good approximation of the
correct potential in all the relevant domain, where the quantum charge
density is different from zero. 

In order to make the algorithm more stable and readily compatible
with PBC, the Fourier series coefficients of the potential can be
written in an auxiliary-function formalism, i.e. as $\mathbf{g}$-dependent
coefficients which correct the analytical Fourier transform coefficients:
\begin{multline}
G\left(\mathbf{g}\right)=\tilde{G}\left(\mathbf{g}\right)+G\left(\mathbf{g}\right)-\tilde{G}\left(\mathbf{g}\right)\\
=\tilde{G}\left(\mathbf{g}\right)+\Delta G\left(\mathbf{g}\right)
\end{multline}

For short-ranged potentials, for the reasons discussed above, one
has that $\Delta G^{short}\left(\mathbf{g}\right)=0$. On the other
hand, for the Coulomb potential one needs to compute the long-range
correction in reciprocal space numerically using fast Fourier transforms.
Once the values of $\Delta G\left(\mathbf{g}\right)$ are known, the
first-image form of the electrostatic potential of the system can
be easily obtained as
\begin{multline}
\hat{v}\left[\rho\right]\left(\mathbf{r}\right)=\sum_{\mathbf{g}}\frac{4\pi}{V}\rho\left(\mathbf{g}\right)\left(\frac{1-\delta_{\mathbf{g}\mathbf{0}}}{g^{2}}+\Delta G\left(\mathbf{g}\right)\right)e^{i\mathbf{g}\cdot\mathbf{r}}\\
=\mathtt{v}\left[\rho\right]\left(\mathbf{r}\right)+\Delta\hat{v}\left[\rho\right]\left(\mathbf{r}\right)
\end{multline}

where the Kronecker $\delta_{\mathbf{g}\mathbf{0}}$ is 1 for $\mathbf{g}=\mathbf{0}$
and 0 otherwise, where 
\begin{equation}
\Delta G\left(\mathbf{0}\right)=\lim_{\mathbf{g}\rightarrow\mathbf{0}}\left(\tilde{G}\left(\mathbf{g}\right)-\frac{1}{g^{2}}\right),
\end{equation}
and where the corrective potential is now computed in reciprocal space
as 
\begin{equation}
\Delta\hat{v}\left[\rho\right]\left(\mathbf{r}\right)=\sum_{\mathbf{g}}\frac{4\pi}{V}\rho\left(\mathbf{g}\right)\Delta G\left(\mathbf{g}\right)e^{i\mathbf{g}\cdot\mathbf{r}}.
\end{equation}
The coefficients entering the calculation of the potential are only
dependent on the geometry of the cell and on the type of potential
that is computed (depending, e.g., whether the whole Coulomb potential
is computed or just its long-range part). Thus, these coefficients
can be computed once and for all at the beginning of a calculation
and the overall computational cost of the procedure becomes negligible.
This is not the case for real-spaces approaches, where a new potential
needs to be computed during the SCF cycle following the change in
the multipole moments of the charge distribution, even though it is
usually not necessary to update it at each SCF step \cite{dabo_prb_2008,dabo_prb_2008_erratum}.
On the other hand, real-space approaches are in principle able to
provide a good approximation of the exact potential profile over the
entire cell and can usually adopt smaller cell-sizes compared to MT
approaches, where the imposed periodicity can significantly alter
the potential at the cell boundaries. 

Reciprocal space approaches are particularly suited for calculations
in the presence of a continuum dielectric medium. In particular, by
extending the use of the auxiliary function coefficients computed
for the potential also to the calculation of the gradient of the potential
\begin{equation}
\nabla\hat{v}\left[\rho\right]\left(\mathbf{r}\right)=\sum_{\mathbf{g}}\frac{4\pi}{V}\rho\left(\mathbf{g}\right)i\mathbf{g}\left(\frac{1-\delta_{\mathbf{g}\mathbf{0}}}{g^{2}}+\Delta G\left(\mathbf{g}\right)\right)e^{i\mathbf{g}\cdot\mathbf{r}}
\end{equation}
it is straightforward to compute the ideal polarization charge by
using Eq. \eqref{eq:rhopol_isol}. All the standard SCCS equations
can then be used straightforwardly, but particular care has to be
given to the calculation of the forces. Indeed, the inter-atomic forces
can be computed from the Hellman-Feynman theorem as 
\begin{equation}
\mathbf{f}_{a}=-\frac{dE\left[\rho^{solute}\right]}{d\mathbf{R}_{a}}=-\int\rho^{solute}\frac{\partial v\left[\rho^{ions}\right]\left(\mathbf{r}\right)}{\partial\mathbf{R}_{a}}\mbox{d}\mathbf{r},
\end{equation}
while the Martyna-Tuckerman correction to the potential introduces
the following contribution 
\begin{multline}
\Delta\mathbf{f}_{a}=-\int\rho^{solute}\frac{\partial\Delta\hat{v}\left[\rho^{ions}\right]\left(\mathbf{r}\right)}{\partial\mathbf{R}_{a}}\mbox{d}\mathbf{r}\\
=-\sum_{\mathbf{g}}\rho^{solute}\left(\mathbf{g}\right)\frac{\partial}{\partial\mathbf{R}_{a}}\left[\frac{4\pi}{V}\left(\sum_{b}^{N^{ions}}z_{b}e^{i\mathbf{g}\cdot\mathbf{R}_{b}}\right)\Delta G\left(\mathbf{g}\right)\right]\\
=-\frac{4\pi z_{a}}{V}\sum_{\mathbf{g}}i\mathbf{g}\rho^{solute}\left(\mathbf{g}\right)e^{i\mathbf{g}\cdot\mathbf{R}_{a}}\Delta G\left(\mathbf{g}\right).
\end{multline}
Similarly, since the forces in the SCCS framework require an additional
term due to the solvent polarization density
\begin{equation}
f_{a,i}^{pol}=-\frac{\partial E^{pol}\left[\rho^{solute},\rho^{pol}\right]}{\partial R_{a,i}}=-\int\rho^{pol}\frac{\partial\Delta\hat{v}\left[\rho^{ions}\right]\left(\mathbf{r}\right)}{\partial R_{a,i}}\mbox{d}\mathbf{r}
\end{equation}
 the proper MT correction needs to be included:
\begin{equation}
\Delta\mathbf{f}_{a}^{pol}=-\frac{4\pi z_{a}}{V}\sum_{\mathbf{g}}i\mathbf{g}\rho^{pol}\left(\mathbf{g}\right)e^{i\mathbf{g}\cdot\mathbf{R}_{a}}\Delta G\left(\mathbf{g}\right).\label{eq:deltaf_mt+sccs}
\end{equation}

\section{Results}

\subsection{Numerical details}

The methods reported in the previous section have been implemented
in a development version of the open-source Quantum-ESPRESSO distribution
\cite{QUANTUM-espresso}. 

Calculations of 0D systems are performed on a pyridine molecule, using
the local-density approximation (LDA) of DFT with a wavefunction cutoff
of 50 Ry. The Brillouin zone is sampled only at the Gamma point. Norm-conserving
pseudo-potentials from the 0.2.2 version of the library of Dal Corso
et al. \cite{pslibrary} are adopted. 

The analysis of 2D systems is performed for a CO molecule adsorbed
in the atop geometry on a fcc Pt (111) surface. In order to speed
up the computational cost for the calculation, a simplified structure
is adopted for the slab, composed by only two layers of metal atoms
in a $\sqrt{3}\times2$ super-cell, with a lattice constant of $2.828$
\AA. Marzari-Vanderbilt \cite{marzarivanderbilt_prl_1999} cold smearing
with a smearing width of 0.03 Ry is used, the Brillouin zone is sampled
with a shifted $4\times4\times1$ reciprocal-space integration grid,
ultrasoft pseudo-potentials and the LDA are adopted, with wavefunction
and density cutoffs of 35 and 280 Ry, respectively. 

For both 0D and 2D systems the accuracy of the methods is tested by
comparing the Hellmann-Feynman forces against the ones obtained by
finite differences of the energy (a displacement step of 0.01 a.u.
is adequate for all the systems studied).

\subsection{Isolated (0D) systems}

\begin{figure}
\includegraphics[width=0.9\columnwidth]{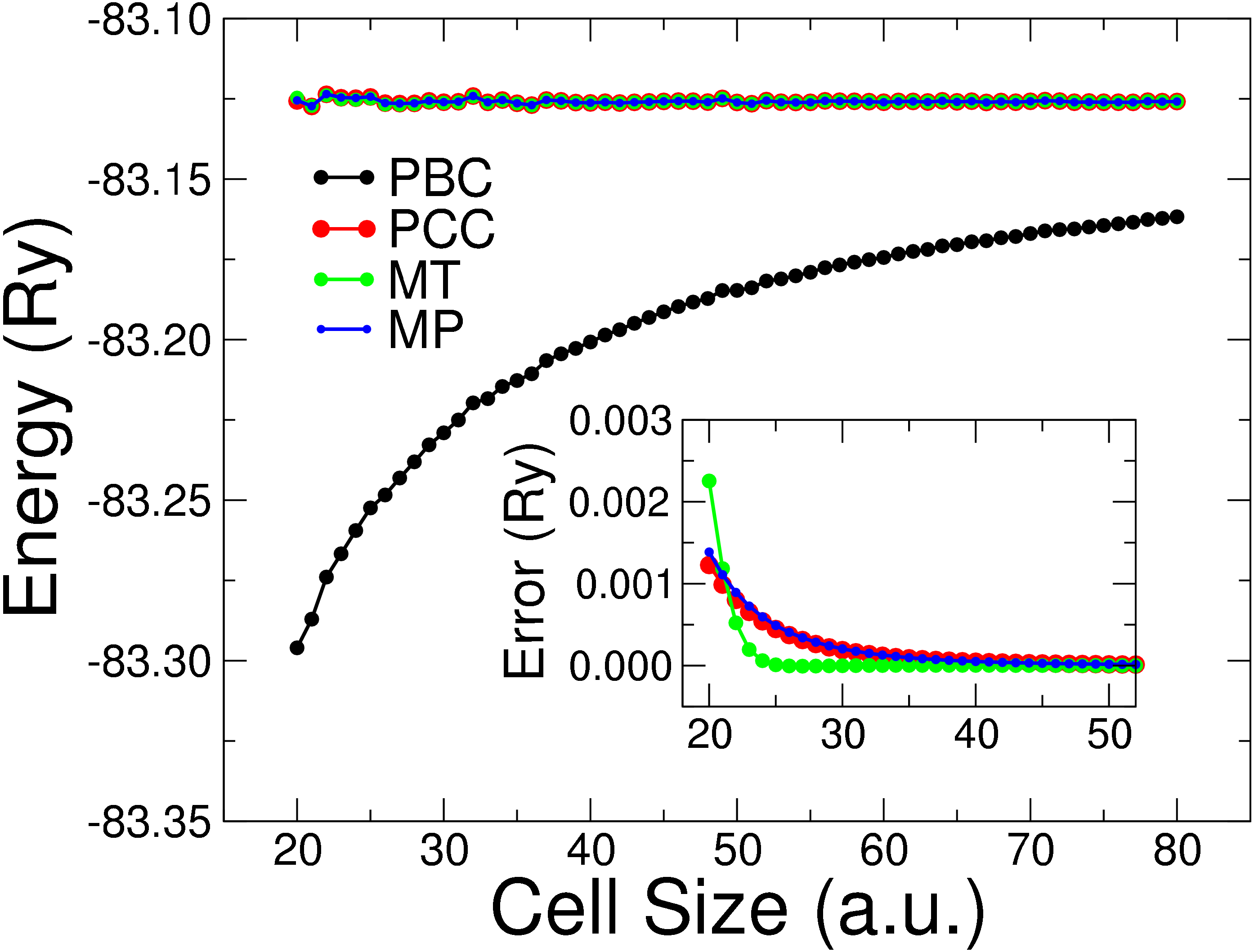}\caption{\label{fig:e_vs_cell_vacuum}Total energy of a pyridine cation in
vacuum as a function of cell size, for PBC calculations and for the
three correction schemes analyzed: Makov-Payne (MP, in blue), Martyna-Tuckerman
(MT, in green), and Point-Countercharge (PCC in red).}
\end{figure}
\begin{figure}
\includegraphics[width=0.9\columnwidth]{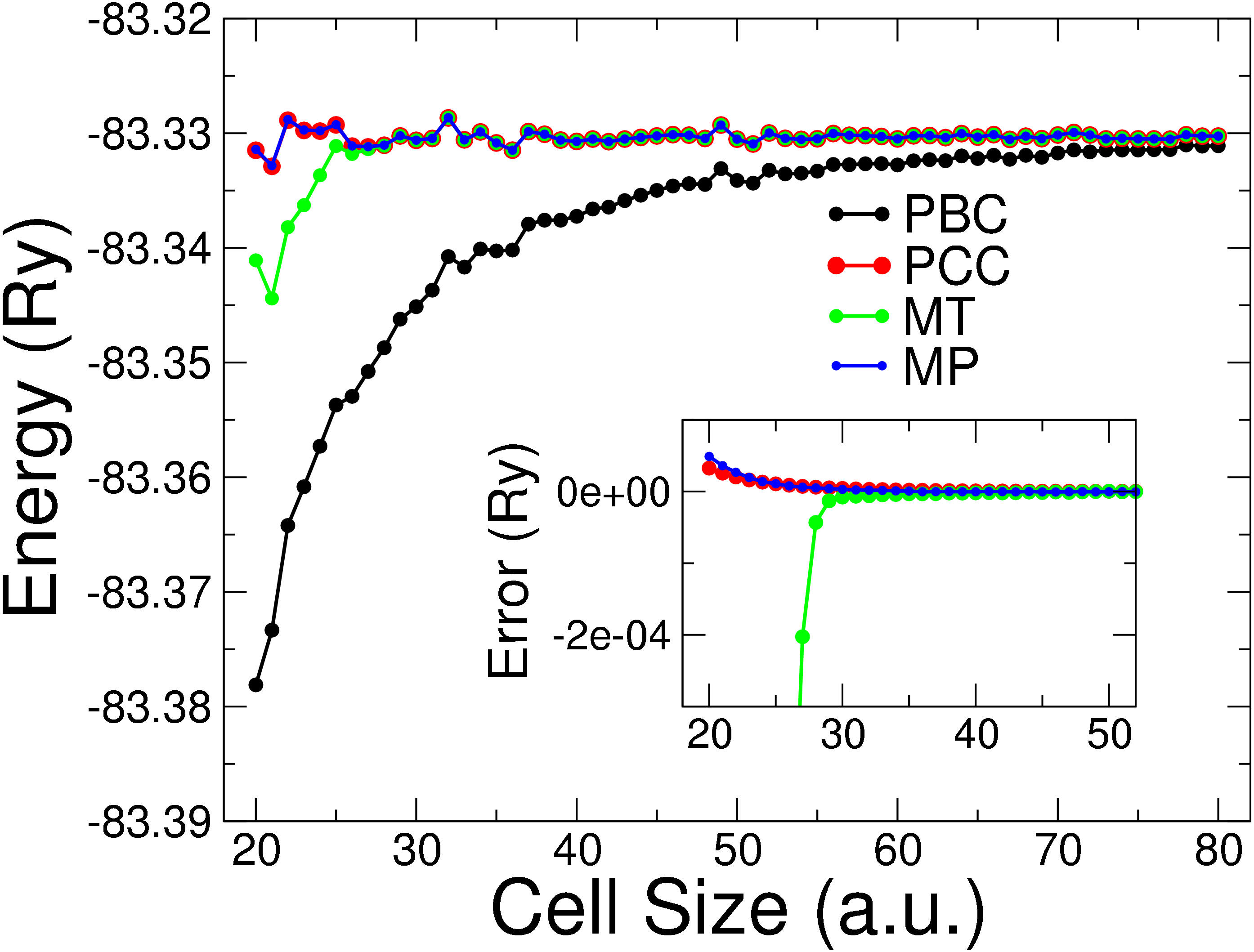}\caption{\label{fig:e_vs_cell_solvent}Total energy of a pyridine cation in
a dielectric medium as a function of cell size, for PBC calculations
and for the three correction schemes analyzed: Makov-Payne (MP, in
blue), Martyna-Tuckerman (MT, in green), and Point-Countercharge (PCC
in red). For the dielectric medium the SCCS parameters optimized to
reproduce aqueous solvation of organic compounds, as derived in Ref.
\cite{andreussi_jcp_2012}, have been used, but only the electrostatic
contribution has been explicitly considered.}
\end{figure}

In Figures \ref{fig:e_vs_cell_vacuum} and \ref{fig:e_vs_cell_solvent}
we report the behavior of the energy of a pyridine cation as a function
of the cell size, for a system in periodic boundary conditions and
for the different correction schemes presented in the Sections above,
both without (Figure \ref{fig:e_vs_cell_vacuum}) and with (Figure
\ref{fig:e_vs_cell_solvent}) a continuum solvent as described by
the SCCS method. As expected, both for the molecule in vacuum and
for the solvated one, the periodic energy decays as the inverse power
of the cell size, with a much less marked dependence for the solvated
cases, due to the dielectric which screens the total charge and dipole
moment of the system. Corrected results from the different methods
are converged for cell sizes of $23$ ($27$) a.u. for the vacuum
(solvent) case, reflecting the larger size of the solvated system.
The MP and PCC energies are found to converge as fast as $L^{-5}$
(as seen in the log-log plot of the residual error, see Figure \ref{fig:mp_contributions_vs_cell}),
while the Martyna-Tuckerman energies become constant and exact for
cell sizes larger than $30$ a.u. On the other hand, while for small
cells the energies computed with the real space approaches and the
Makov-Payne are still less than 1 mRy away from the converged results,
the Martyna-Tuckerman approach shows significant errors, even exceeding
the uncorrected periodic energy. The same trend is reflected in the
calculation of the electrostatic contribution to solvation free energies.

It is important to note that Makov-Payne energies are almost identical
to the ones obtained with the self-consistent real space approach.
This validates the hypothesis, assumed by Makov and Payne, that the
polarization of the charge density of the system due to periodic images
affects only marginally its energy. The same behavior is true for
solvated systems, with the Makov-Payne and the PCC methods almost
exactly on top of each other.

\begin{figure}
\includegraphics[width=0.9\columnwidth]{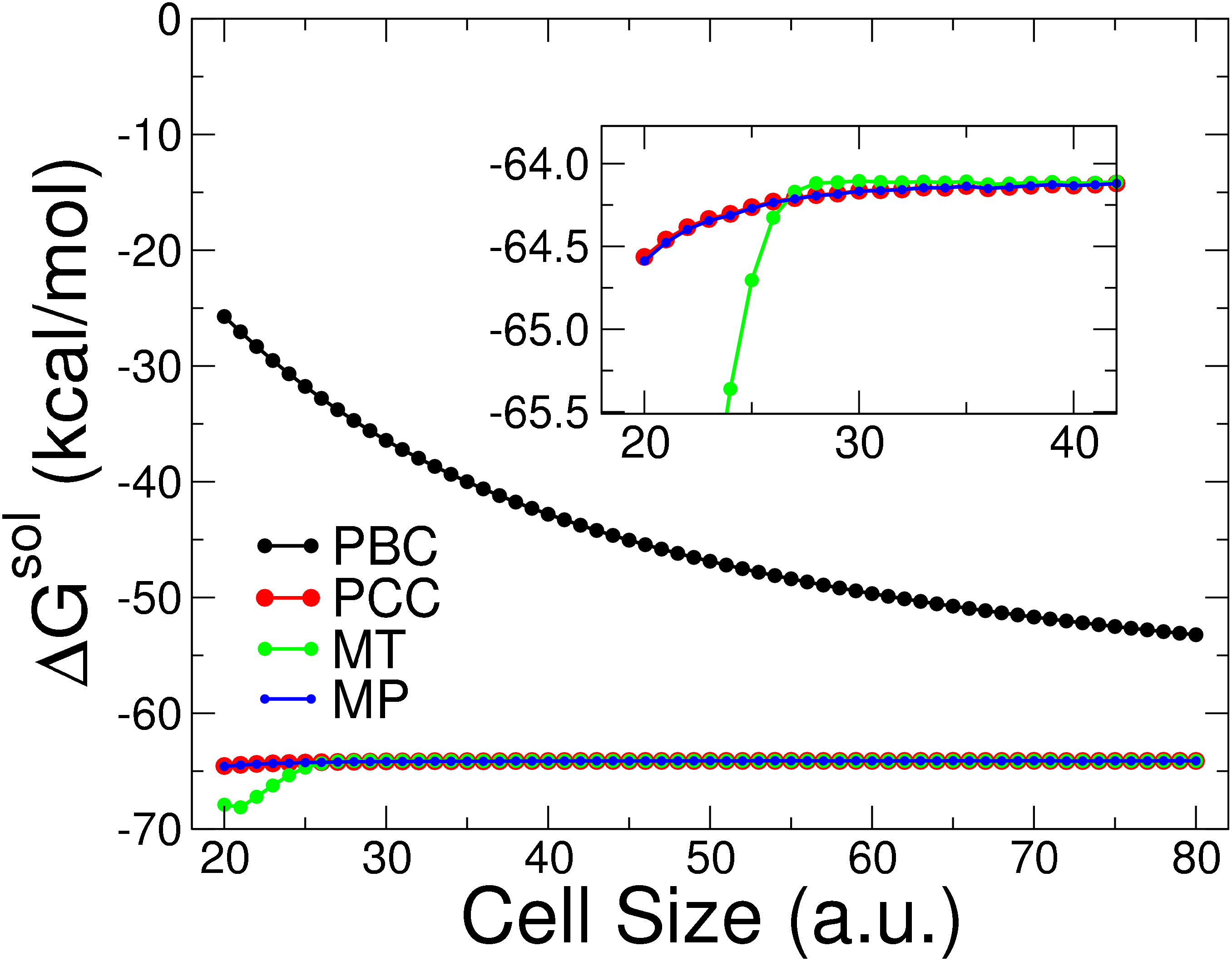}\caption{\label{fig:dgsol_vs_cell}Solvation energies of a pyridine cation
in continuum dielectric medium as a function of cell size, for PBC
calculations and for the three correction schemes analyzed: Makov-Payne
(MP, in blue), Martyna-Tuckerman (MT, in green), and Point-Countercharge
(PCC in red). For the dielectric medium the SCCS parameters optimized
to reproduce aqueous solvation of organic compounds, as derived in
Ref. \cite{andreussi_jcp_2012}, have been used, but only the electrostatic
contribution has been explicitly considered.}
\end{figure}

Electrostatic solvation free energies, computed as the difference
in total energy between the solvated and the vacuum case, are reported
in Figure \ref{fig:dgsol_vs_cell} and substantially reflect what
was found above: MP and PCC calculations give well converged results
(errors smaller than 0.5 kcal/mol) for all the system sizes considered,
while the MT scheme gives large errors up to cell sizes of $26$ a.u.

\begin{figure}
\includegraphics[width=0.9\columnwidth]{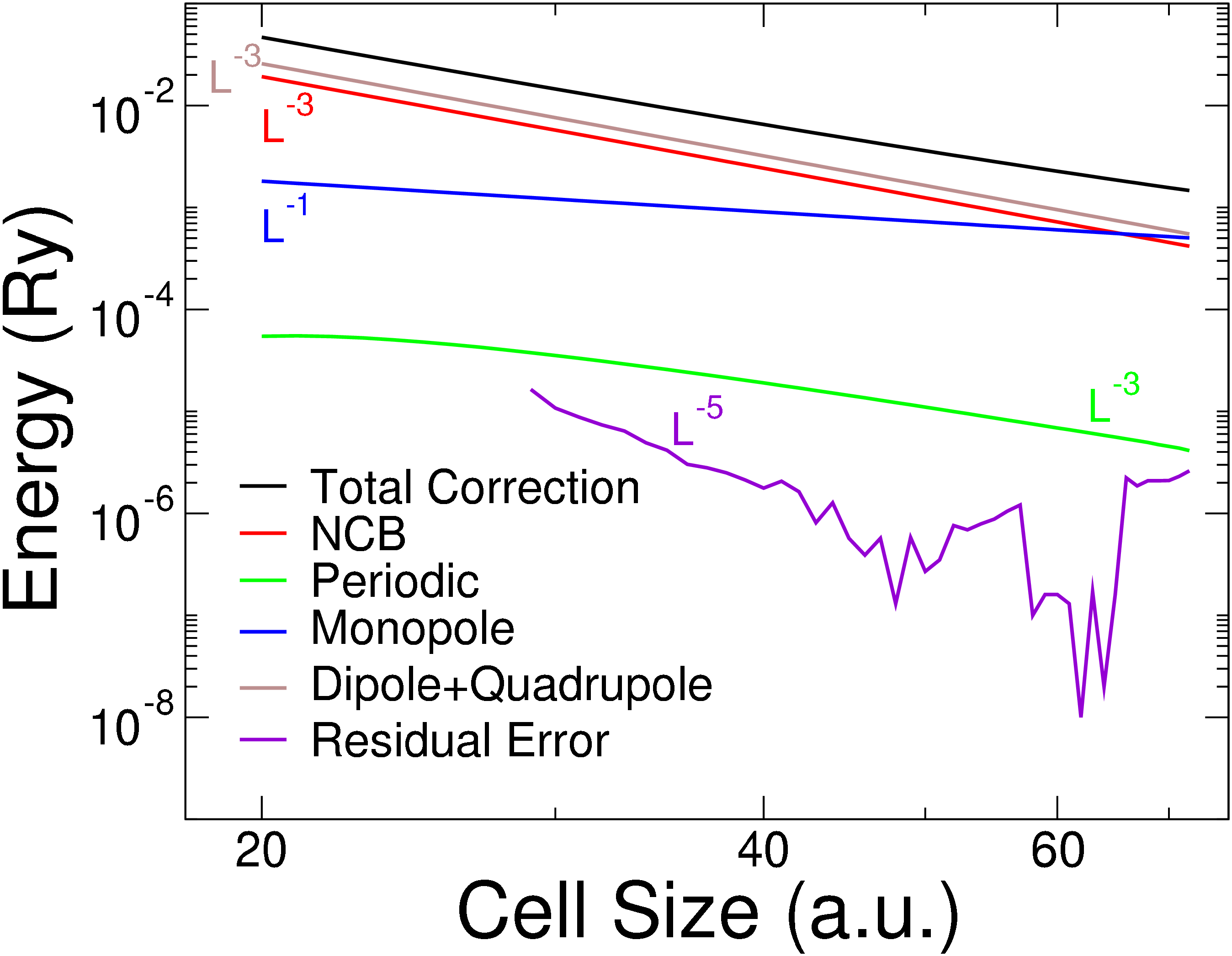}\caption{\label{fig:mp_contributions_vs_cell}Cell-size dependence of Makov-Payne
(Monopole and Dipole+Quadrupole) and polarization specific (NCB and
Periodic) contributions to the energy. Reported deacy exponents are
estimated from fitting the large cell-size part of the figure. The
residual error, after the different contributions have been subtracted
from the total energy of the system, is shown to decay faster than
$L^{-3}$ for cell sizes up to 40 a.u., while being negligible for
larger cell sizes.}
\end{figure}

When looking at the different contributions to the Makov-Payne correction
for a solvated system (Figure \ref{fig:mp_contributions_vs_cell}),
it appears that all contribute significantly. In particular, the effects
of the periodic images on polarizing the dielectric close to the solute
are important, especially considering that, although small, such polarization
is not charge-neutral in the case of charged solutes. 

\begin{figure}
\includegraphics[width=0.9\columnwidth]{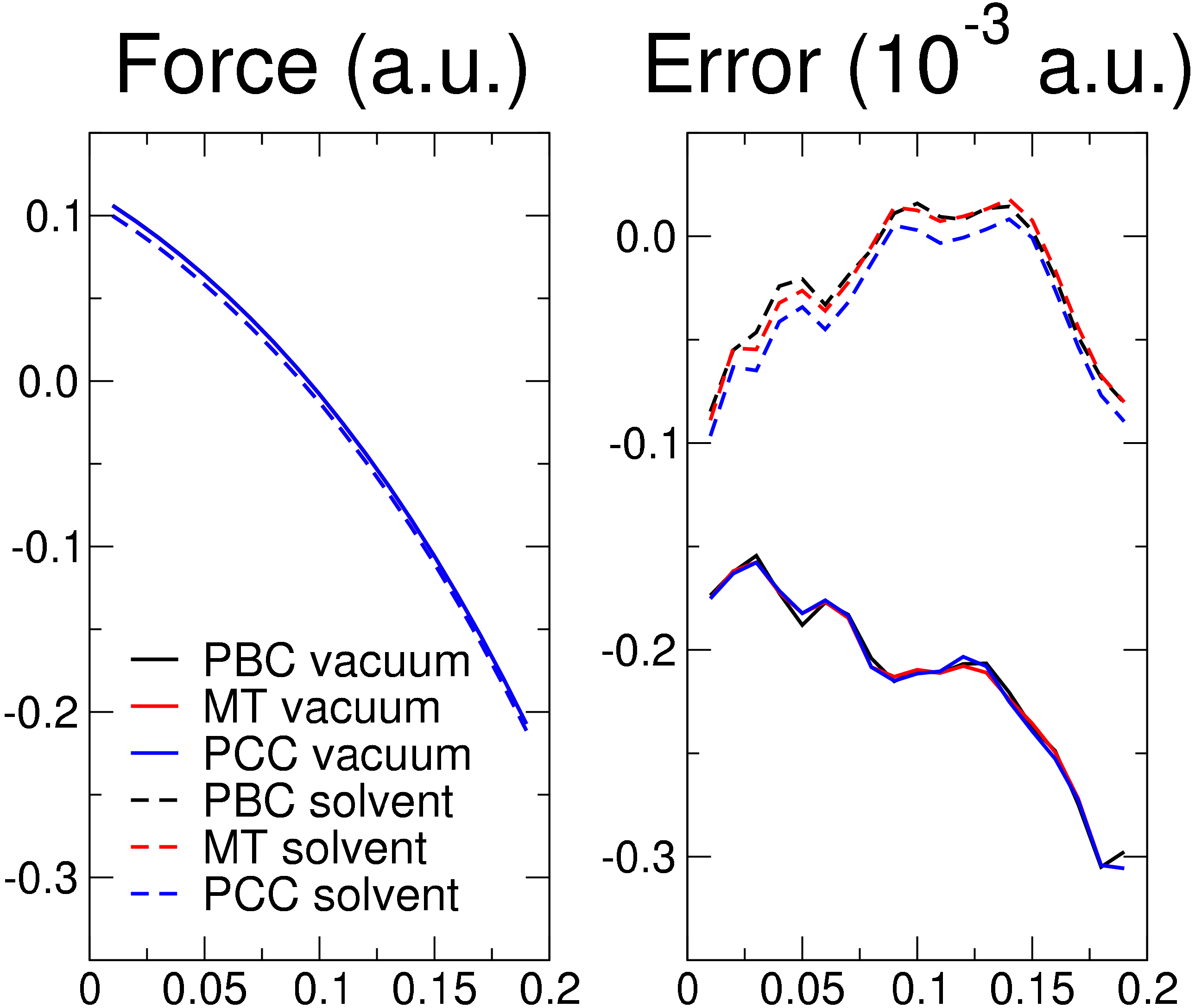}\caption{\label{fig:0dforce_and_error_vs_dx}Hellmann-Feynman force (left panel)
and error on forces (right panel) computed via finite differences
for the nitrogen atom of the pyridine cation, in vacuum and in a continuum
dielectric medium, with PBC or with MT and PCC correction schemes.}
\end{figure}

In order to validate the formulas derived and the implementation of
the different methods, the errors on the analytic forces have been
reported in Figure \ref{fig:0dforce_and_error_vs_dx} for the different
approaches considered as well as for the fully periodic case, with
and without a continuum solvent. All the different methods show a
very similar behavior, with errors almost three orders of magnitude
smaller than the absolute value of the computed force. It is important
to stress that, even though the reported errors are not negligible,
all the methods are in agreement to what is found for the periodic
calculations without the solvent, which represents the internal benchmark
of this work. Lower values for these errors can be obtained by fine
tuning the setup of the calculations.

\subsection{Slab (2D) geometries}

\begin{figure}
\includegraphics[width=0.9\columnwidth]{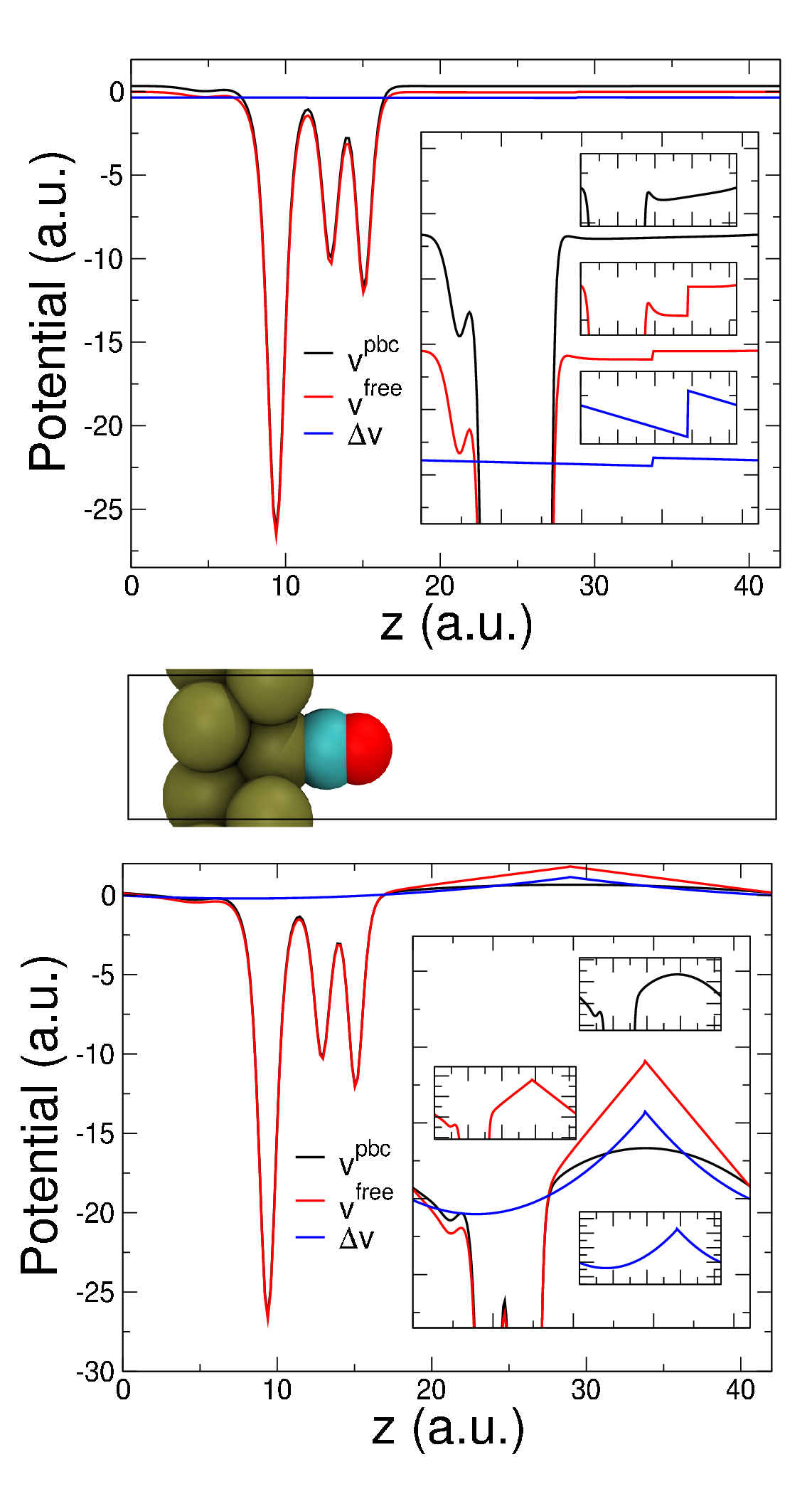}\caption{\label{fig:v_vs_z_vacuum}Electrostatic potential along the axis passing
through the C-O bond for a neutral (top) and charged (bottom) slab
in vacuum, as represented in the middle panel (Pt atoms in yellow,
carbon atom in light blue, oxygen atom in red). The PBC potential
($v^{pbc}$, in black), PCC correction ($\Delta v$, in blue) and
corrected potential ($v^{free}$, in red) are reported and compared
in the main panels and in the insets.}
\end{figure}
\begin{figure}
\includegraphics[width=0.9\columnwidth]{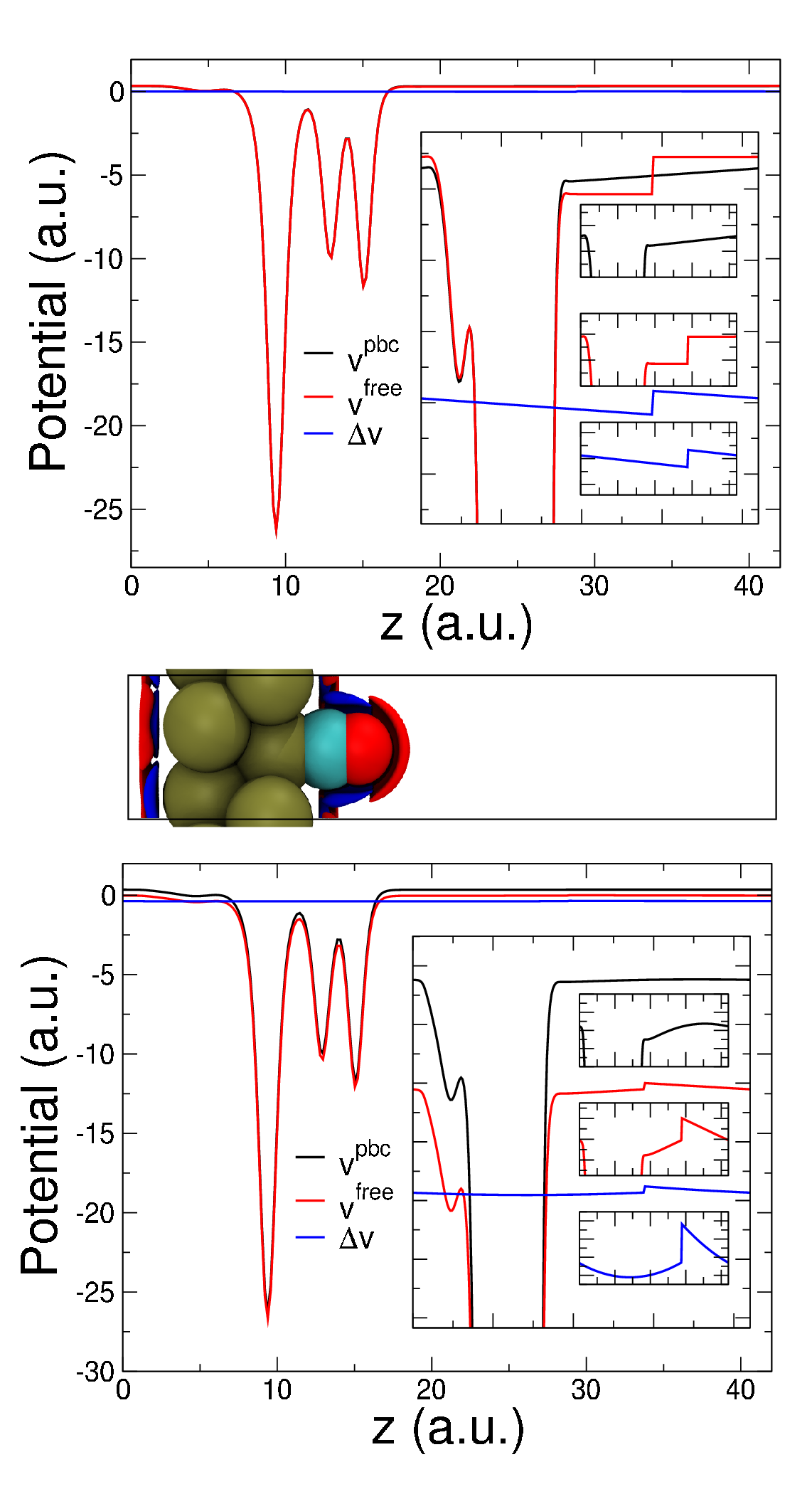}\caption{\label{fig:v_vs_z_solvent}Electrostatic potential along the axis
passing through the C-O bond for a neutral (top) and charged (bottom)
slab in a continuum dielectric, as represented in the middle panel
(Pt atoms in yellow, carbon atom in light blue, oxygen atom in red,
positive and negative polarization charges are visualized as red and
blue solid isosurfaces). The PBC potential ($v^{pbc}$, in black),
PCC correction ($\Delta v$, in blue) and corrected potential ($v^{free}$,
in red) are reported and compared in the main panels and in the insets.}
\end{figure}
\begin{figure}
\includegraphics[width=0.95\columnwidth]{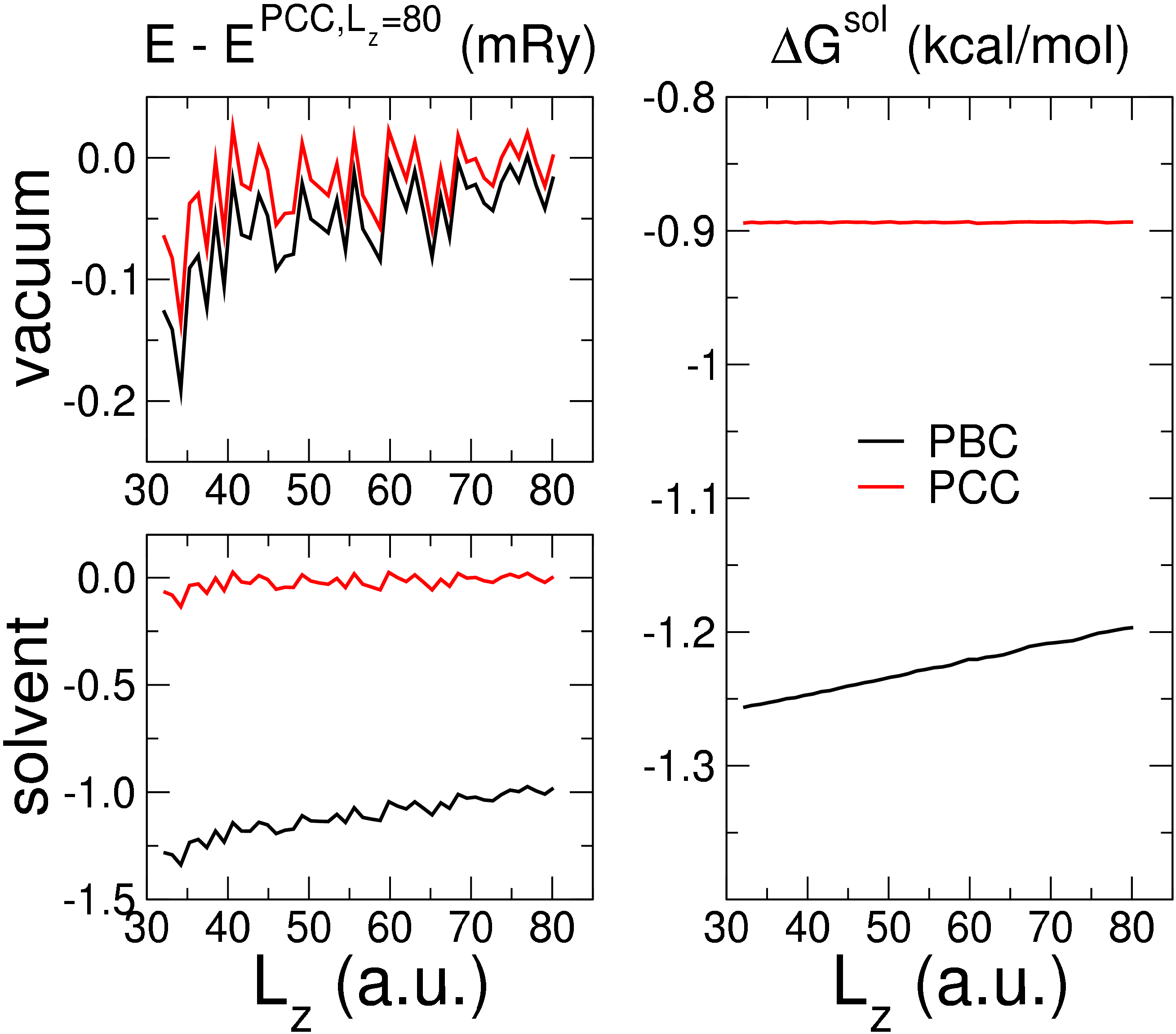}

\caption{\label{fig:e_and_dgsol_vs_cell}Total energy in vacuum (top left)
and in solution (bottom left), and solvation energy (right) of a neutral
slab as a function of cell thickness. For the dielectric medium the
SCCS parameters optimized to reproduce aqueous solvation of organic
compounds, as derived in Ref. \cite{andreussi_jcp_2012}, have been
used, but only the electrostatic contribution has been explicitly
considered.}
\end{figure}

When considering neutral two dimensional systems, the presence of
a component of the dipole moment along the axis normal to the plane
of the slab will create a step in the electrostatic potential. Due
to periodic boundary conditions requiring the potential to be continuous
at the boundary of the simulation cell, a deformation of the potential
in the entire cell will take place (see top panels of Figures \ref{fig:v_vs_z_vacuum}
and \ref{fig:v_vs_z_solvent}). When looking at the energy of the
neutral system as a function of cell size, Figure \ref{fig:e_and_dgsol_vs_cell},
it looks that periodic boundary conditions have very different effects
in vacuum and in solution. While the correction amounts to only a
fraction of mRy for the system in vacuum, the effects of periodicity
is more than ten times larger in solvated systems. The same trends
are clearly reflected in the behavior of the solvation energy of the
system, which shows PBC errors of \textasciitilde{}0.3 kcal/mol even
for the largest cells considered. In fact, the overall error seems
to decay very slowly with cell thickness. This behavior is due to
the spurious finite electric field in the interfacial region where
the dielectric properties of the environment change: given the expression
of the polarization charge (Eq. \ref{eq:rhopol_isol}), the presence
of the artificial linear electrostatic potential induces a substantial
amount of polarization in the continuum environment. Thus, even though
the dielectric medium should compensate the intrinsic dipole of the
slab, the interaction with the finite field due to periodic boundary
conditions over-stabilizes larger polarization charges and increases
the PBC artifacts on the energy of solvated 2D systems. The simple
2D PCC correction proposed in this article is enough to remove this
artifact and to provide energies in solutions which are well behaved
with respect to the size of the simulation cell, as clearly shown
in Figure \ref{fig:e_and_dgsol_vs_cell}. The corrected form of the
potential (red lines in Figures \ref{fig:v_vs_z_vacuum} and \ref{fig:v_vs_z_solvent})
is almost constant in the regions of space above and below the slab,
thus providing the right contribution to the polarization of the surrounding
medium. 

In the case of a charged two-dimensional system, a converged reference
value for the energy of the system is not accessible, since the field,
and thus the electrostatic energy density, of a planar distribution
of charge is constant in space. Thus, the electrostatic potential
and the total energy will linearly increase with the size of the cell
axis perpendicular to the plane of the slab.

Such a behaviour is correctly recovered in the trends of the potential
in Figures \ref{fig:v_vs_z_vacuum} and \ref{fig:v_vs_z_solvent}
for the system in vacuum and in solution, respectively. The presence
of the dielectric medium significantly screens the electrostatic potential,
which shows much smaller variations with respect to the case in vacuum.

\begin{figure}
\includegraphics[width=0.9\columnwidth]{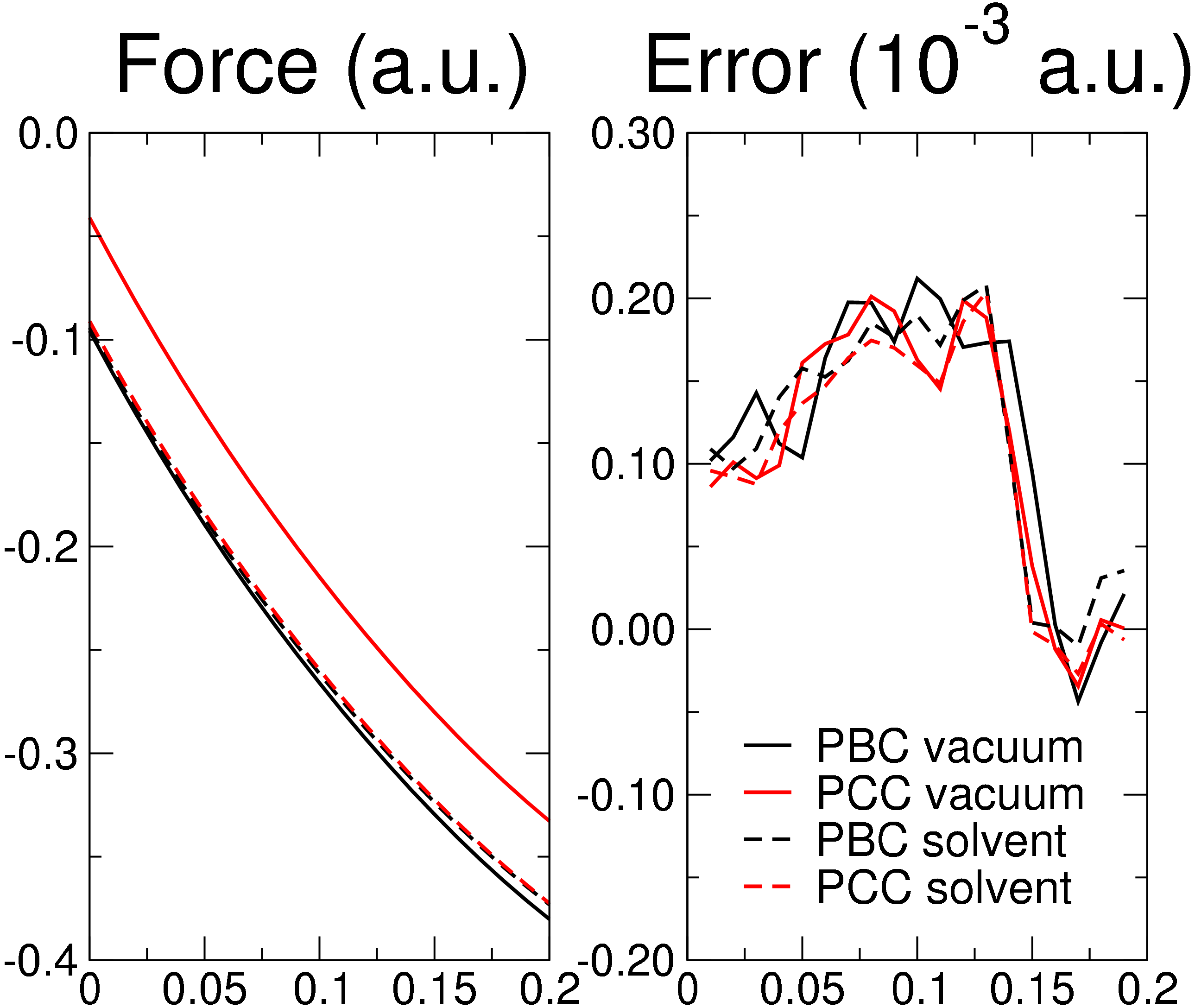}\caption{\label{fig:2dforce_and_error_vs_dx}Hellmann-Feynman force (left panel)
and error on forces (right panel) computed via finite differences
for the carbon atom in the charged slab calculation, in vacuum and
in a continuum dielectric medium, with and without PBC.}

\end{figure}

As in the case of the isolated 0D system, the analytical forces computed
with the PBC correction, both with and without the continuum solvent,
show errors of similar magnitude to the ones obtained from calculations
without the correction (see Figure \ref{fig:2dforce_and_error_vs_dx}).
Self-consistent convergence of the polarization charge for the calculation
of the charged slab in periodic boundary conditions appears to be
hindered by the presence of the NCB density. In particular, a very
large number of iterations are required in order to achieve the same
accuracy on the polarization charge as in other calculations (i.e.
with mean squared variation of the polarization density of the order
of $10^{-12}$a.u.). Even though the polarization density due to the
neutralizing charge background has been shown to add a spurious cell-size-dependent
term to the energy, it is important to note that in order to properly
describe forces in a charged slab, the NCB polarization needs to be
taken into account explicitly.

\section{Conclusions}

To conclude, an extension to three current methods (Makov-Payne \cite{makov_prb_1995},
Martyna-Tuckerman \cite{martyna_jcp_1999}, and PCC \cite{dabo_prb_2008})
to correct for periodic-boundary conditions in systems of reduced
dimensionality is presented, that allows to treat quantum systems
immersed in a continuum dielectric. Two different geometries have
been explicitly addressed, namely isolated (0D) and slab (2D) configurations.
The modified Makov-Payne correction is summarized by Eq. \ref{eq:makov-payne_sccs},
where the energy of an isolated system solvated in a continuum dielectric
is expressed in terms of its artificially periodic counterpart plus
a post-processing, computationally inexpensive, correction term. The
main results for the PCC scheme are, instead, summarized by Eq. \ref{eq:deltav_pcc+sccs_0d}
and \ref{eq:deltav_pcc+sccs_2d} for the 0D and 2D cases, respectively,
where the real-space corrections to the electrostatic potential of
the solvated system are reported. While the Martyna-Tuckerman scheme
is shown to be intrinsically more easy to adapt to the electrostatic
equations defined by the SCCS approach, we underline the derivation
of the correction's contributions to interatomic forces in the presence
of a continuum dielectric, as summarized in Eq. \ref{eq:deltaf_mt+sccs}. 

The analytical modifications introduced due to the presence of the
solvent have been shown to be accurate in all the different methods.
The derivation of the modifications proposed here can be easily extended
to more complex methods, such as the density countercharge, or to
linear (1D) systems. Results are shown to converge reasonably fast
with system size. A comparison of the different approaches indicates
that the Martyna-Tuckerman is the technique that converges faster
with cell size, but can produce erratic results for cell sizes that
are too small. The behavior of Makov-Payne and PCC correction schemes
are very similar and are much smoother with system size, so that smaller
cells with respect to MT can be used at the price of a small loss
in accuracy. The very good agreement of the forces, computed analytically
and via finite-differences of the energy, is a further test on all
derivations and their implementation.

\subsection*{Appendix A: Ionic charge distribution in SCCS and real space periodic
boundary corrections.}

As discussed in Ref \cite{andreussi_jcp_2012}, the way the ionic
charge density is described in the SCCS has no effect on the computed
polarization density or on the polarization energy, provided that
the ionic charge is well within the range where the dielectric constant
is exactly one. Since Gaussian charges of fixed spread are used to
model nuclei through the SCCS calculation, solvation energies were
shown to be independent of their spread for a large range of values
(Figure (18) of Ref. \cite{andreussi_jcp_2012}). Nonetheless, when
computing the correction for periodic boundary conditions in real
space, either in the self-consistent (PCC) of in the non self-consistent
(Makov-Payne) case, it is important that the ionic cores are treated
on the same footing as the calculation of the ionic electrostatic
energy. If Ewald summations are used to model the ionic contribution
to the total energy, ions need to be considered as point-like in the
calculation of the system's multipole moments $d^{solute}$ and $Q^{solute}$
that enter in Eq. (\eqref{eq:makov-payne_vacuum}). On the other hand,
in order for the polarization energy to be independent of the shape
of the ionic density, a consistent description need to be used in
Eq. (\eqref{eq:energy_pol_isol}) and (\eqref{eq:makov-payne_sccs}). 

Different choices of ionic shapes will in general give rise to different
multipole moments. Nonetheless, if we chose to work in a coordinate
frame originating from the center of ionic charges, the nuclei's contribution
to the system's dipole vanishes regardless of the fact that they be
described as point-like or Gaussian charges. Thus, with such a choice
of origin, only the quadrupole moment of the system depends on the
shape of the ionic charge distribution. In order to correctly remove
all the terms depending on the quadrupole, Gaussian-shaped ions have
to be used when correcting the solvation energy, while point-like
nuclei have to be used for the remaining part. Thus an extra term
of the form
\begin{multline}
\Delta E^{0D,Gaussian}=\\
\frac{\pi}{3L^{3}}q^{pol}\left(Q^{solute,Gaussian}-Q^{solute,point-like}\right)
\end{multline}
for 0D systems, or 
\begin{multline}
\Delta E^{2D,Gaussian}=\\
\frac{\pi}{AL_{z}}q^{pol}\left(Q_{zz}^{solute,Gaussian}-Q_{zz}^{solute,point-like}\right)
\end{multline}
for 2D systems has to be added to the correction of the energy. An
extra correction on the forces is also needed in the case of 0D systems
(while it vanishes due to symmetry in the 2D case), namely 
\begin{equation}
\Delta f_{a}^{0D,Gaussian}=-\frac{2\pi}{3L^{3}}q^{pol}\left(\frac{z_{a}\sigma_{a}}{\sqrt{\pi}}\right),
\end{equation}
where $z_{a}$ and $\sigma_{a}$ are the atomic charge and spread
of atom $a$.

\section*{Acknowledgments}
\begin{acknowledgments}
The authors acknowledge Celine Dupont, Ismaila Dabo and Stefano de
Gironcoli for useful discussions and the Swiss Platform for Advanced
Scientific Computing (PASC) for funding.
\end{acknowledgments}

\bibliographystyle{apsrev}

\end{document}